\documentclass[oldversion]{aa}  
\usepackage{graphicx,amsmath}
\usepackage{amssymb}
\usepackage{color}
\usepackage{natbib}		
\usepackage{url}
\usepackage{multirow}
\usepackage{threeparttable}
\bibpunct{(}{)}{;}{a}{}{,} 
\usepackage{txfonts}
\usepackage{subfig}

\begin{document}

   \title{Hubble PanCET: An extended upper atmosphere of neutral hydrogen around the warm Neptune GJ\,3470 b}

   \author{
   V.~Bourrier\inst{1},     
   A.~Lecavelier des Etangs\inst{2},     
   D.~Ehrenreich\inst{1},    
   J.~Sanz-Forcada\inst{3},   
   R.~Allart\inst{1},     
   G.E.~Ballester\inst{4} ,   
   L.A.~Buchhave\inst{5},    
   O.~Cohen\inst{6},   
   D.~Deming\inst{7}    
   T.M.~Evans\inst{8},   
   A.~Garc\'ia Mu$\tilde{\mathrm{n}}$oz\inst{9},   
   G.W.~Henry\inst{10},   
   T.~Kataria\inst{11},   
   P.~Lavvas\inst{12},   
   N.~Lewis\inst{13},    
   M.~L\'opez-Morales\inst{14},     
   M.~Marley\inst{15},    
   D.K.~Sing\inst{16,17},    
   H.R.Wakeford\inst{18},   
	}

\authorrunning{V.~Bourrier et al.}
\titlerunning{Lyman-$\alpha$ transit observations of GJ\,3470 b}

\offprints{V.B. (\email{vincent.bourrier@unige.ch})}

\institute{
Observatoire de l'Universit\'e de Gen\`eve, 51 chemin des Maillettes, 1290 Sauverny, Switzerland.  
\and
Institut d'Astrophysique de Paris, CNRS, UMR 7095 \& Sorbonne Universit\'es, UPMC Paris 6, 98 bis bd Arago, 75014 Paris, France 
\and
Centro de Astrobiologia (CSIC-INTA), ESAC Campus, P.O. Box 78, E-28691 Villanueva de la Canada, Madrid, Spain   
\and
Dept. of Planetary Sciences \& Lunar \& Planetary Lab., University of Arizona, 1541 E Univ. Blvd., Tucson, AZ 85721, USA 
\and
DTU Space, National Space Institute, Technical University of Denmark, Elektrovej 328, DK-2800 Kgs. Lyngby, Denmark      
\and
Lowell Center for Space Science and Technology, University of Massachusetts, Lowell, Massachusetts 01854, USA.	
\and
Department of Astronomy, University of Maryland, College Park, Maryland 20742, USA.  
\and
Astrophysics Group, Physics Building, University of Exeter, Stocker Road, Exeter EX4 4QL, UK   
\and
Zentrum f\"{u}r Astronomie und Astrophysik, Technische Universit\"{a}t Berlin, D-10623 Berlin, Germany    
\and
Center of Excellence in Information Systems, Tennessee State University, Nashville, TN 37209, USA    
\and
NASA Jet Propulsion Laboratory, 4800 Oak Grove Dr, Pasadena, CA 91109, USA   
\and
Groupe de Spectroscopie Mol\'eculaire et Atmosph\'erique, Universit\'e de Reims, Champagne-Ardenne, CNRS UMR 7331, France    
\and
Department of Astronomy and Carl Sagan Institute, Cornell University, 122 Sciences Drive, 14853, Ithaca, NY
\and
Harvard-Smithsonian Center for Astrophysics, 60 Garden Street, Cambridge, MA 02138, USA   
\and
NASA Ames Research Center, Moffett Field, California, USA    
\and
Department of Physics and Astronomy, University of Exeter, Exeter, UK.   
\and
Department of Earth and Planetary Sciences, Johns Hopkins University, Baltimore, MD, USA.    
\and
Space Telescope Science Institute, 3700 San Martin Drive, Baltimore, MD 21218, USA
}

   \date{} 
 
  \abstract
{GJ\,3470b is a warm Neptune transiting a M dwarf star at the edge of the evaporation desert. It offers the possibility to investigate how low-mass, close-in exoplanets evolve under the irradiation from their host stars. We observed three transits of GJ\,3470b in the Lyman-$\alpha$ line with the Hubble Space Telescope (HST) as part of the Panchromatic Comparative Exoplanet Treasury (PanCET) program. Absorption signatures are detected with similar properties in all three independent epochs, with absorption depths of 35$\pm$7\% in the blue wing of the line, and 23$\pm$5\% in the red wing. The repeatability of these signatures, their phasing with the planet transit, and the radial velocity of the absorbing gas, allow us to conclude that there is an extended upper atmosphere of neutral hydrogen around GJ\,3470 b. We determine from our observations the stellar radiation pressure and XUV irradiation from GJ\,3470 and use them to perform numerical simulations of GJ\,3470b's upper atmosphere with the EVaporating Exoplanets (EVE) code. The unusual redshifted signature can be explained by the damping wings of dense layers of neutral hydrogen extending beyond the Roche lobe and elongated in the direction of the planet motion. This structure could correspond to a shocked layer of planetary material formed by the collision of the expanding thermosphere with the wind of the star. The blueshifted signature is well explained by neutral hydrogen atoms escaping at rates of about 10$^{10}$\,g\,s$^{-1}$, blown away from the star by its strong radiation pressure, and quickly photoionized, resulting in a smaller exosphere than that of the warm Neptune GJ\,436b. The stronger escape from GJ\,3470b, however, may have led to the loss of about 4--35\% of its current mass over its $\sim$2\,Gyr lifetime. 
 }


   \maketitle

\section{Introduction}
\label{intro} 

Extended exospheres were first detected around strongly irradiated gas giants through transit observations in the ultraviolet spectrum of their host stars. The hot Jupiters HD\,209458b (\citealt{VM2003,VM2004,VM2008}; \citealt{Ehrenreich2008}; \citealt{BJ_Hosseini2010}) and HD\,189733b (\citealt{Lecav2010,Lecav2012}; \citealt{Bourrier2013}) revealed Lyman-$\alpha$ transit absorption depths about ten times larger than their planetary transit at optical wavelengths. Absorption signatures measured in the blue wing of the Lyman-$\alpha$ line traced the presence of neutral hydrogen gas beyond the gravitational influence of the planets and moving away from the stars at velocities reaching $\sim$100 -- 200\,km\,s$^{-1}$. This evaporation arises from the deposition of stellar X-ray and extreme ultraviolet radiation (XUV) at the base of the planet thermosphere, leading to its hydrodynamic expansion up to altitudes of several planetary radii (e.g. \citealt{VM2003}; \citealt{Lammer2003}; \citealt{Lecav2004}; \citealt{GarciaMunoz2007}; \citealt{Johnstone2015}, \citealt{Guo2016}).\\

\begin{figure}     
\includegraphics[trim=0cm 0cm 0cm 0cm,clip=true,width=\columnwidth]{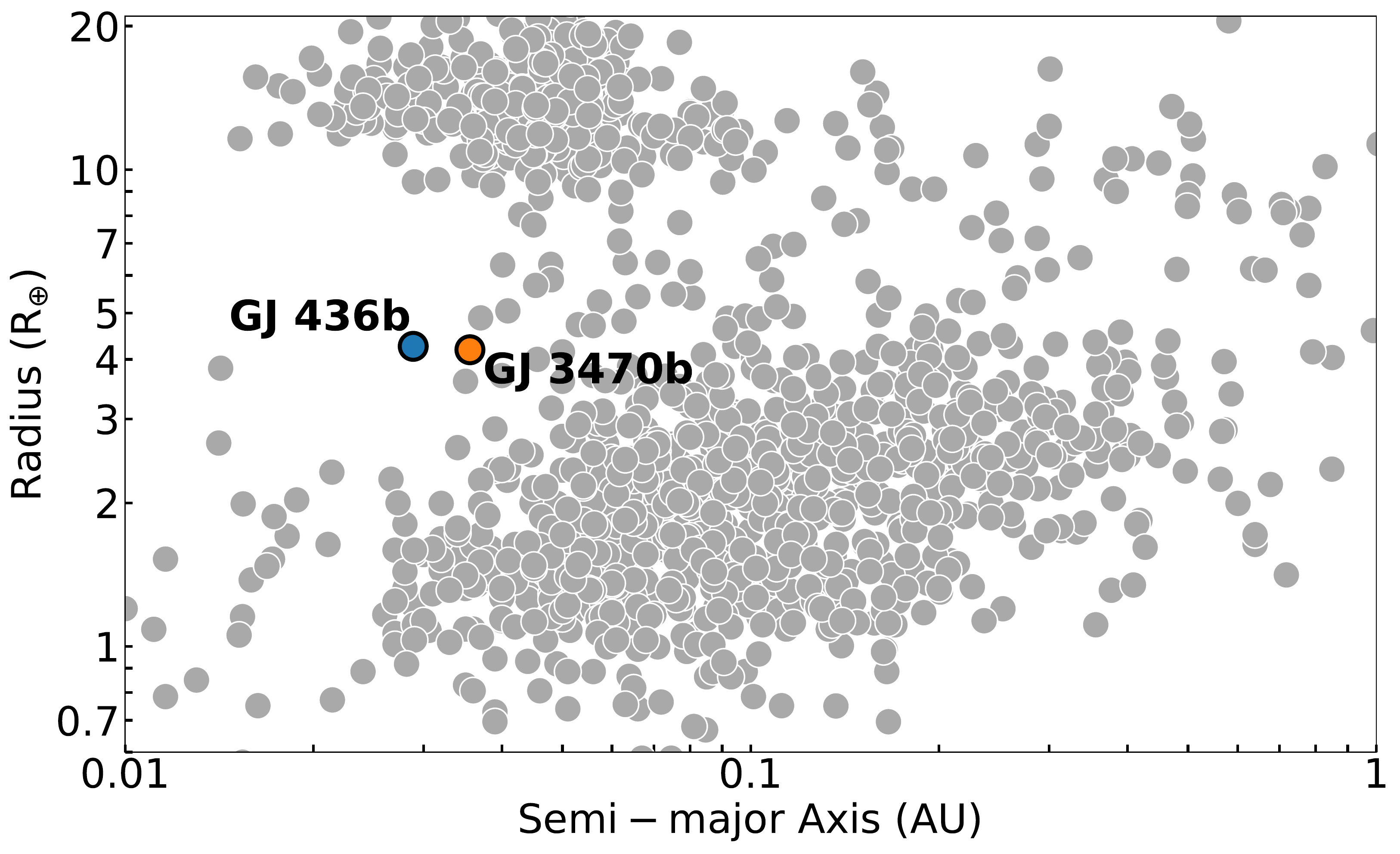}
\caption[]{Distribution of planetary radius as a function of distance to the star. Physical properties of exoplanets were extracted from the Extrasolar Planets Encyclopaedia (exoplanet.eu) in April 2018. Evaporation shaped the population of hot close-in planets, likely resulting in the formation of a desert of hot planets with intermediate size. Warm Neptunes GJ\,3470b and GJ\,436b stands at the edge of this desert.}
\label{fig:R_a_fig}
\end{figure}

Despite tremendous mass losses on the order of thousands of tons per second, hot Jupiters are massive enough that evaporation only weakly affects their long-term evolution (e.g. \citealt{Yelle2004}; \citealt{Lecav2007}; \citealt{Hubbard2007}; \citealt{Ehrenreich_desert2011}). There exists, however, a dearth of exoplanets at small orbital distances (Fig.~\ref{fig:R_a_fig}) ranging from sub-Jupiters (\citealt{Lecav2007}; \citealt{Davis2009}; \citealt{szabo_kiss2011}; \citealt{Mazeh2016}) to super-Earths (\citealt{beauge2013}; \citealt{SanchisOjeda2014}; \citealt{Lundkvist2016}; \citealt{Fulton2017,Fulton2018}). Theoretical studies suggest that atmospheric escape plays a major role in shaping this so-called desert (e.g. \citealt{Lopez2012,Lopez2013}; \citealt{Jin2014}; \citealt{Kurokawa2014}) and valley (\citealt{Owen2017}; \citealt{Jin2018}) in the exoplanet population, with strongly irradiated, low-density planets unable to retain their escaping gaseous atmosphere. Some of the very hot rocky super-Earths could thus be the evaporated remnants of more massive planets (``chthonian'' planets, \citealt{Lecav2004}).\\
 
In this context, it is essential to observe exoplanets at the borders of the desert of hot, intermediate-size planets (Fig.~\ref{fig:R_a_fig}) to understand its origin and how planetary and stellar properties influence atmospheric escape and the evolution of close-in exoplanets. The case of the warm Neptune GJ\,436b, which revealed extremely deep transit signatures in the UV (\citealt{Kulow2014}; \citealt{Ehrenreich2015}; \citealt{Lavie2017}), suggests that members of this class of exoplanets are ideal targets to be characterized through their upper atmosphere. The mild irradiation of these warm Neptunes, frequently found around late-type stars, is sufficient to lead to their evaporation, but the combination of stellar wind interactions with the low radiation pressure and low photoionization allows their neutral hydrogen exosphere to extend up to tens of planetary radii (\citealt{Bourrier2015, Bourrier2016}, \citealt{Lavie2017}). GJ 436b has a similar radius than the warm Neptune GJ\,3470b, but a density two times lower. Both planets orbit at similar distance (in stellar radius) from M dwarfs, but GJ3470b is hosted by an earlier-type star. Together, these systems thus offer the possibility to investigate how stellar irradiation and planetary properties affect the upper atmosphere and evolution of warm Neptunes.\\

Following-up on the study of GJ\,436b, we obtained Lyman-$\alpha$ observations of GJ\,3470b as part of the HST Panchromatic Comparative Exoplanet Treasury program (PanCET, GO 14767, P.I. Sing \& Lopez-Morales), which targets 20 exoplanets for the first large-scale simultaneous UltraViolet, Optical, Infrared (UVOIR) comparative study of exoplanets. GJ\,3470b was discovered through radial velocity and transit observations by \citet{Bonfils2012}. It has a low bulk density (0.8\,g\,cm$^{-3}$, \citealt{Awiphan2016}) indicative of a hydrogen-helium rich atmosphere (\citealt{Demory2013_GJ3470b}, \citealt{Chen_2017}). With a period of $\sim$3.3 days and a moderately bright and nearby M1.5 dwarf host star (V = 12.3 , d=29.5\,pc), GJ\,3470b is a favorable target for atmospheric characterization through transit spectroscopy. Its near-infrared transmission spectrum shows no atmospheric features at the precision obtainable with HST/WFC3 in staring mode (\citealt{Ehrenreich2014}) or from the ground (\citealt{Crossfield2013}). In the optical, several studies point to a tentative increase in the planet radius towards short wavelengths (\citealt{Fukui2013}; \citealt{Nascimbeni2013}, \citealt{Biddle2014}; \citealt{Dragomir2015}; \citealt{Awiphan2016}; \citealt{Chen_2017}). As in HD\,189733b, HD\,209458b, WASP-6b, and HAT-P-12b (\citealt{Lecav2008a, Lecav2008b}; \citealt{Nikolov2015}; \citealt{Sing2016}), this slope could be attributed to Rayleigh scattering in an extended atmosphere of molecular hydrogen and helium. The low gravitational potential of GJ\,3470b and the strong high-energy irradiation from its host star suggests that it sports an extended upper atmosphere (\citealt{Salz2016b}). Here we report on the first transit observations of GJ\,3470b in the \ion{H}{i} Lyman-$\alpha$ line, aimed at detecting its extended exosphere of neutral hydrogen. We present the observations in Sect.~\ref{sec:obs_datared} and describe their analysis in Sect.~\ref{sec:tr_ana}. We derive from our observations the intrinsic Lyman-$\alpha$ line of GJ\,3470 in Sect.~\ref{sec:Ly_rec} and its XUV spectrum in Sect.~\ref{sec:em_lines}. The reconstructed stellar spectrum is used in Sect.~\ref{sec:descr_EVE} to interpret the Lyman-$\alpha$ transit observations of GJ\,3470b with our 3D numerical model of evaporating atmosphere. We discuss our results in Sect.~\ref{sec:discuss}.

\begin{table*}[tbh]
\caption{Physical properties of the GJ\,3470 system.}                                                 
\begin{threeparttable}
\begin{tabular}{llccccc}
\hline
\hline
\noalign{\smallskip}
Parameters        & Symbol      & Value          & Reference       \\
\noalign{\smallskip}
\hline
\noalign{\smallskip}
Distance from Earth     & $D_{\mathrm{*}}$    &    29.45$\pm$0.07\,pc &    \\
\noalign{\smallskip}
Star radius             & $R_{\mathrm{*}}$    &    0.547$\pm$0.018$\,R_{\mathrm{\sun}}$ & \citealt{Awiphan2016}    \\
\noalign{\smallskip}
Star mass            & $M_{\mathrm{*}}$      &    0.539$\stackrel{+0.047}{_{-0.043}}$ $\,M_{\mathrm{\sun}}$  & \citealt{Demory2013_GJ3470b}  \\
\noalign{\smallskip}
Star rotation period            & $P_{\mathrm{*}}$      &    20.70$\pm$0.15\,d  & \citealt{Biddle2014}  \\
\noalign{\smallskip}
Star age            & $\tau_{\mathrm{*}}$      &    $\sim$2\,Gyr  & This work  \\
\noalign{\smallskip}
Heliocentric stellar radial velocity            & $\gamma_{\mathrm{*}/\mathrm{\sun}}$      &   26.51691$\pm$5.3$\times$10$^{-4}$\,km\,s$^{-1}$  & \citealt{Bonfils2012}  \\
\noalign{\smallskip}
Planet radius (358\,nm)     & $R_{\mathrm{p}}$    &  4.8$\pm$0.2$\,R_{\mathrm{Earth}}$   & \citealt{Chen_2017}      \\\noalign{\smallskip}
Planet mass     & $M_{\mathrm{p}}$    &   13.9$\pm$1.5$\,M_{\mathrm{Earth}}$ & \citealt{Demory2013_GJ3470b}   \\
\noalign{\smallskip}
Orbital period        & $P_{\mathrm{p}}$   &  3.33665173$\pm$5.9$\times$10$^{-7}$\,days   & \citealt{Chen_2017}    \\
\noalign{\smallskip}
Transit center  & $T_{\mathrm{0}}$                       &   2455983.70417$\pm$0.00011$\,BJD_\mathrm{TDB}$  & \citealt{Chen_2017} \\
\noalign{\smallskip}
Semi-major axis   & $a_{\mathrm{p}}$  & 0.0355$\pm$0.0019$\,au$ & \citealt{Awiphan2016} \\
\noalign{\smallskip}
Eccentricity       & $e$         &    0.086$\stackrel{+0.050}{_{-0.036}}$ &       \\
\noalign{\smallskip}
Argument of periastron  & $\omega$       &    -123.5$\stackrel{+21.9}{_{-44.4}}^{\circ}$  &      \\
\noalign{\smallskip}
Inclination       & $i_{\mathrm{p}}$           &  89.13$\stackrel{+0.26}{_{-0.34}}^{\circ}$   & \citealt{Awiphan2016} \\
\noalign{\smallskip}
\hline
\hline
\end{tabular}
\begin{tablenotes}[para,flushleft]
Notes: The distance from GJ\,3470 was derived from its GAIA parallax measurement (33.96$\pm$0.06\,mas; \citealt{Gaia2016,Gaia2018}). The values for $e$ and $\omega$ were derived from radial velocity measurements of GJ\,3470 (X. Bonfils, private communication). 
  \end{tablenotes}
  \end{threeparttable}
\label{tab:param_sys}
\end{table*}


\section{Observations and data reduction}
\label{sec:obs_datared}

\subsection{HST STIS observations}
\label{sec:obs}

We observed the M dwarf GJ\,3470 with the Space Telescope Imaging Spectrograph (STIS) instrument on board the Hubble Space Telescope (HST). We used the STIS/G140M grating (spectral range 1195 to 1248\,\AA, spectral resolution $\sim$20\,km\,s$^{-1}$) to measure the transit of GJ\,3470b in the stellar Lyman-$\alpha$ line. Three visits of five consecutive HST orbits each were scheduled on 28 November 2017 (Visit A), 4 December 2017 (Visit B), and 7 January 2018 (Visit C). Visits were centered around the planet mid-transit to provide sufficient coverage before and after the transit, and scheduled at slightly different orbital phases to increase the temporal sampling (see e.g. Fig.~\ref{fig:LC_grid}). Scientific exposures lasted 2154\,s in all HST orbits, except during the first orbit that is shorter because of target acquisition (exposure time 1824\,s). Data obtained in time-tagged mode were reduced with the CALSTIS pipeline, which includes the flux and wavelength calibration, and divided in all HST orbits into 6 sub-exposures with duration 304 or 359\,s.\\

\begin{figure}     
\includegraphics[trim=0cm 0cm 0cm 0cm,clip=true,width=\columnwidth]{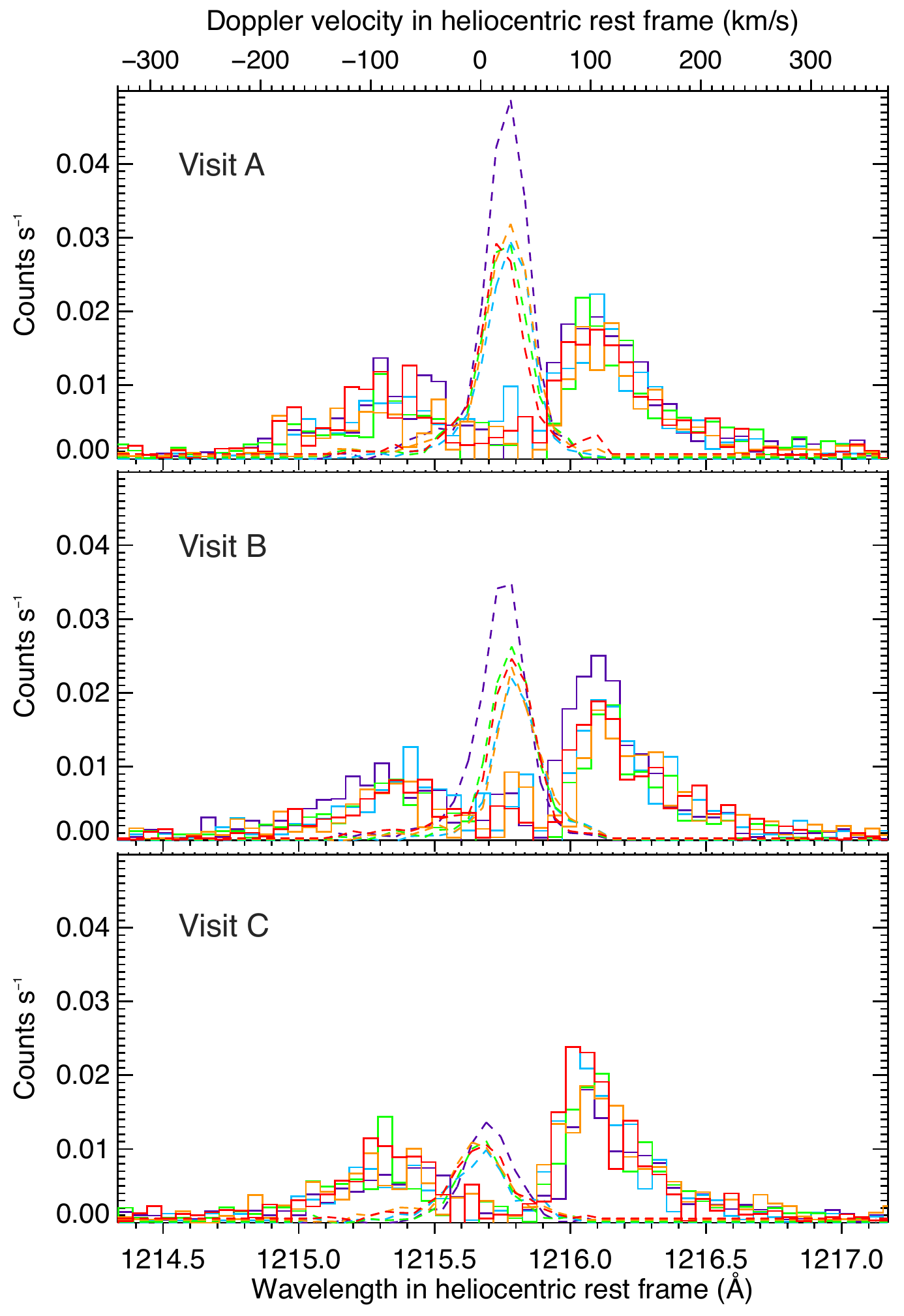}
\caption[]{Raw spectra of GJ\,3470 Lyman-$\alpha$ line (histograms) after correction for the geocoronal emission line (superimposed as a dashed line). Colors correspond to HST consecutive orbits in each visit (orbits 1 to 5 are plotted in purple, blue, green, orange, red).}
\label{fig:airglow}
\end{figure}

Geocoronal airglow emission from the upper atmosphere of Earth (\citealt{VM2003}) contaminates the stellar Lyman-$\alpha$ emission line in the raw data (Fig.~\ref{fig:airglow}). Although CALSTIS corrects the final 1D spectra for airglow contamination, it is recommended to treat with caution the regions where the airglow is stronger than the stellar flux. In these regions, the airglow-corrected flux is usually overestimated (Fig.~\ref{fig:airglow}). The strength and position of the airglow varies with the epoch of observation, affecting different parts of the spectrum. Therefore, we excluded from our analysis the wavelength window 1215.70--1215.86\,\AA\, in Visit A, 1215.70--1215.81\,\AA\, in Visit B, and 1215.61--1215.78\,\AA\, in Visit C, defined in the heliocentric rest frame.

\subsection{Flux calibration}

\begin{figure}     
\includegraphics[trim=0cm 0cm 0cm 0cm,clip=true,width=\columnwidth]{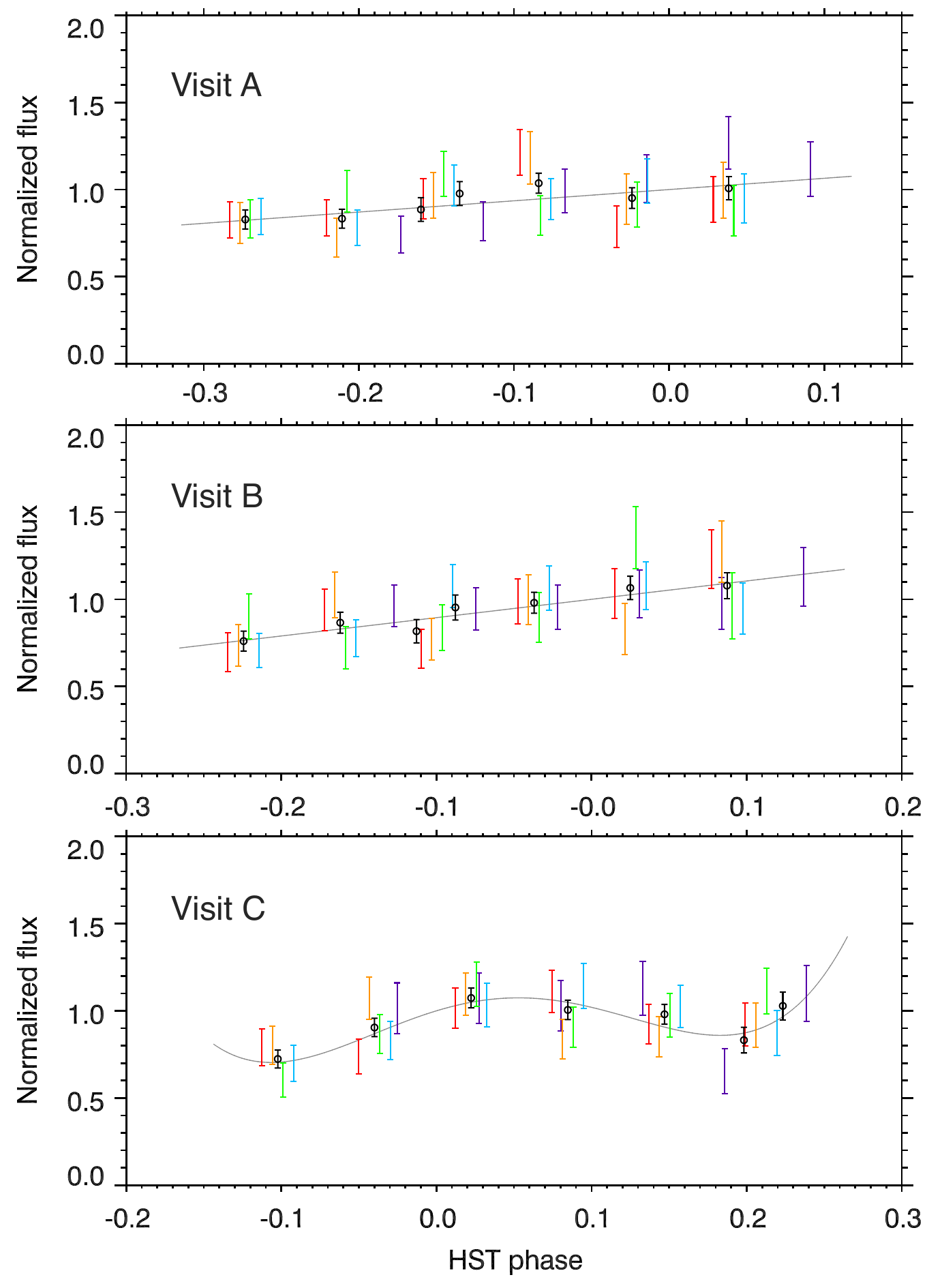}
\caption[]{Lyman-$\alpha$ fluxes for sub-exposures integrated over the entire line and phase-folded on the HST orbital period ($P_\mathrm{HST}$ = 96\,min). Phase is between -0.5 and 0.5. Solid grey lines correspond to the best-fit breathing model to the data, which have been scaled to the same nominal flux for the sake of comparison. Black points show sub-exposures binned manually to highlight the breathing trend. Colors correspond to HST consecutive orbits in each visit (orbits 1 to 5 are plotted in purple, blue, green, orange, red).}
\label{fig:LC_breath}
\end{figure}

STIS observations are affected by variations in the telescope throughput caused by thermal variations that the HST experiences during each orbit (e.g. \citealt{Brown2001}; \citealt{Sing2008a}; \citealt{Huitson2012}). This ``breathing'' effect modifies the flux balance within an HST orbit, and is detected in the three visits (Fig.~\ref{fig:LC_breath}). As expected from previous measurements with the G140M grating, the shape and amplitude of the breathing variations change between visits but the orbit-to-orbit variations within a single visit are both stable and highly repeatable (e.g. \citealt{Bourrier2013}; \citealt{Ehrenreich2015}). Correcting for the breathing variations is particularly important with respect to the first-orbit exposure because it is shorter and shifted to later HST orbital phases compared to subsequent exposures (Fig.~\ref{fig:LC_breath}). As a result, the average flux over the first orbit is subjected to a different bias compared to the other orbits. \\

We fitted a breathing model based on \citet{Bourrier2017_HD976} to the sub-exposure spectra integrated over the entire Lyman-$\alpha$ line (1214.7--1216.7\,\AA\, minus the range contaminated by the airglow). This choice is motivated by the achromaticity of the breathing variations, and by the need for a high SNR to ensure an accurate correction. The breathing was modeled as a polynomial function phased with the period of the HST around the Earth ($P_{\mathrm{HST}}$ = 96\,min). The nominal flux unaffected by the breathing effect was fitted for in each HST orbit, to prevent the overcorrection of putative orbit-to-orbit variations caused by the star or the planet. The model was oversampled in time and averaged within the time window of each exposure before comparison with the data. We used the Bayesian Information Criterion (BIC, \citealt{Liddle2007}) as a merit function to determine the best polynomial degree for the breathing variations. We obtained linear variations in visits A and B, and a 4-th order polynomial function in Visit C (Fig.~\ref{fig:LC_breath}). The spectra in each sub-exposure were then corrected by the value of the best-fit breathing function at the time of mid-exposure. \\

\section{Transit analysis in the Lyman-$\alpha$ line}
\label{sec:tr_ana}

We searched for flux variations between the Lyman-$\alpha$ spectra of GJ\,3470 that would arise from planetary absorption or stellar variability, following the same approach as in previous studies (e.g. \citealt{Bourrier2013, Bourrier2017_HD976}). Spectra were first compared two by two in each visit to identify those showing no significant variations, which could be considered as representative of the quiescent, unocculted stellar line. This was done by searching for all features characterized by flux variations with S/N larger than 3, and extending over more than 3 pixels ($\sim$0.16\,\AA\,, larger than STIS spectral resolution). Then, we coadded these stable spectra to create a reference for the intrinsic stellar line in each visit, which was used to characterize the variable features observed in the other spectra. We caution that noise in the spectra makes it difficult to identify precisely the spectral range of a given variable feature. Nonetheless, signatures consistent with absorption of the quiescent stellar line were identified in each visit. We detail their analysis in the following sections, and we plot in Fig.~\ref{fig:LC_grid} the transit light curves in the spectral bands showing absorption or remaining stable over each visit.\\

\subsection{Visit A}

Lyman-$\alpha$ line spectra are consistent between observations taken outside of the transit (orbits 1, 2, 5). Compared to their average, taken as reference for the intrinsic stellar line, orbits 3 and 4 both revealed absorption signatures in the blue wing and the red wing of the line. These orbits were taken near the ingress and egress of the planetary transit, respectively.
\begin{itemize}
\item \textit{Red wing :} the signatures are detected within a similar spectral range in orbits 3 and 4 (+21 to +60\,km\,s$^{-1}$), with absorption depths of 53.8$\pm$13.0\% and 44.7$\pm$14.1\%, respectively (Fig.~\ref{fig:LC_grid}). 
\item \textit{Blue wing :} the most significant absorption signatures are found at high velocities, within -163 to -124\,km\,s$^{-1}$ (46.7$\pm$14.5\%) in orbit 3, and within -150 to -71\,km\,s$^{-1}$ (43.4$\pm$9.1\%) in orbit 4. However, a visual inspection and the analysis of orbits 3-4 average spectrum suggests that absorption extends to lower velocities. In the range -150 to -32\,km\,s$^{-1}$, we measure consistent absorption depths of 28.8$\pm$9.7\% and 32.8$\pm$9.5\% for orbits 3 and 4, respectively (Fig.~\ref{fig:LC_grid}). \\
\end{itemize}

\subsection{Visit B}
\label{sec:ana_VB}

Lyman-$\alpha$ line spectra are also consistent between out-of-transit orbits 1-2-5 in Visit B, except for a significantly larger flux at the peak of the red wing in orbit 1 (between +61 and +126\,km\,s$^{-1}$). After scaling visits A and B spectra to the same flux level, we found that Visit B spectra in orbits 2 and 5 are in good agreement with Visit A reference spectrum. This suggests that the feature in the first orbit of Visit B is a transient brightening of the quiescent stellar line. We thus built the reference spectrum for Visit B as the average between orbits 1 (excluding the brightened region), 2 and 5. Compared to this reference, we identified absorption signatures in orbits 3 and 4, obtained during the planetary transit and shortly after the egress (Fig.~\ref{fig:LC_grid}).
\begin{itemize}
\item \textit{Red wing :} the strongest signature is detected in orbit 3 (38.9$\pm$11.5\% within +8 to +74\,km\,s$^{-1}$), while orbit 4 shows a marginal signature at larger velocities (26.7$\pm$9.2\% within +34 to +87\,km\,s$^{-1}$).
\item \textit{Blue wing :} the most significant absorption signatures are found within -110 to -18\,km\,s$^{-1}$ (53.3$\pm$14.2\%) in orbit 3, and within -97 to -57\,km\,s$^{-1}$ (61.9$\pm$13.4\%) in orbit 4.  \\
\end{itemize}

\subsection{Visit C}

Visit C revealed short-term variability that makes the identification of quiescent/variable spectra more difficult. The largest variation occurs in the first orbit at the peak of the red wing (+38 to +116\,km\,s$^{-1}$), with the Lyman-$\alpha$ line distorted and significantly fainter than in subsequent orbits. We also observe brightenings in the far red wing in orbit 4, and at the peak of the blue wing in orbit 5. However these latter variations are marginal, and we thus built the reference spectrum for Visit C as the average between orbits 1 (excluding the fainter region), 2 and 5. These orbits were obtained well outside of the planetary transit, close to their counterparts used as reference in visits A and B (Fig.~\ref{fig:LC_grid}). Compared to Visit C reference spectrum, we identified absorption signatures in orbits 3 and 4, obtained at mid-transit and after the egress.

\begin{itemize}
\item \textit{Red wing :} absorption signatures in the red wing are marginal, but interestingly they overlap with the signatures detected in previous visits. The most significant signatures are found within +38 to +77\,km\,s$^{-1}$ (56.7$\pm$14.7\%) in orbit 3, and within +51 to +90\,km\,s$^{-1}$ (25.9$\pm$8.9\%) in orbit 4.
\item \textit{Blue wing :} orbit 3 shows significant absorption between -94 and -41\,km\,s$^{-1}$ (56.7$\pm$14.7\%), whereas we do not detect any significant absorption signature in orbit 4. \\
\end{itemize}

\subsection{Summary}
\label{sec:sum_obs}

\begin{figure*}
\centering
\begin{minipage}[b]{\textwidth}
\includegraphics[trim=0cm 0cm 0cm 0cm,clip=true,width=\columnwidth]{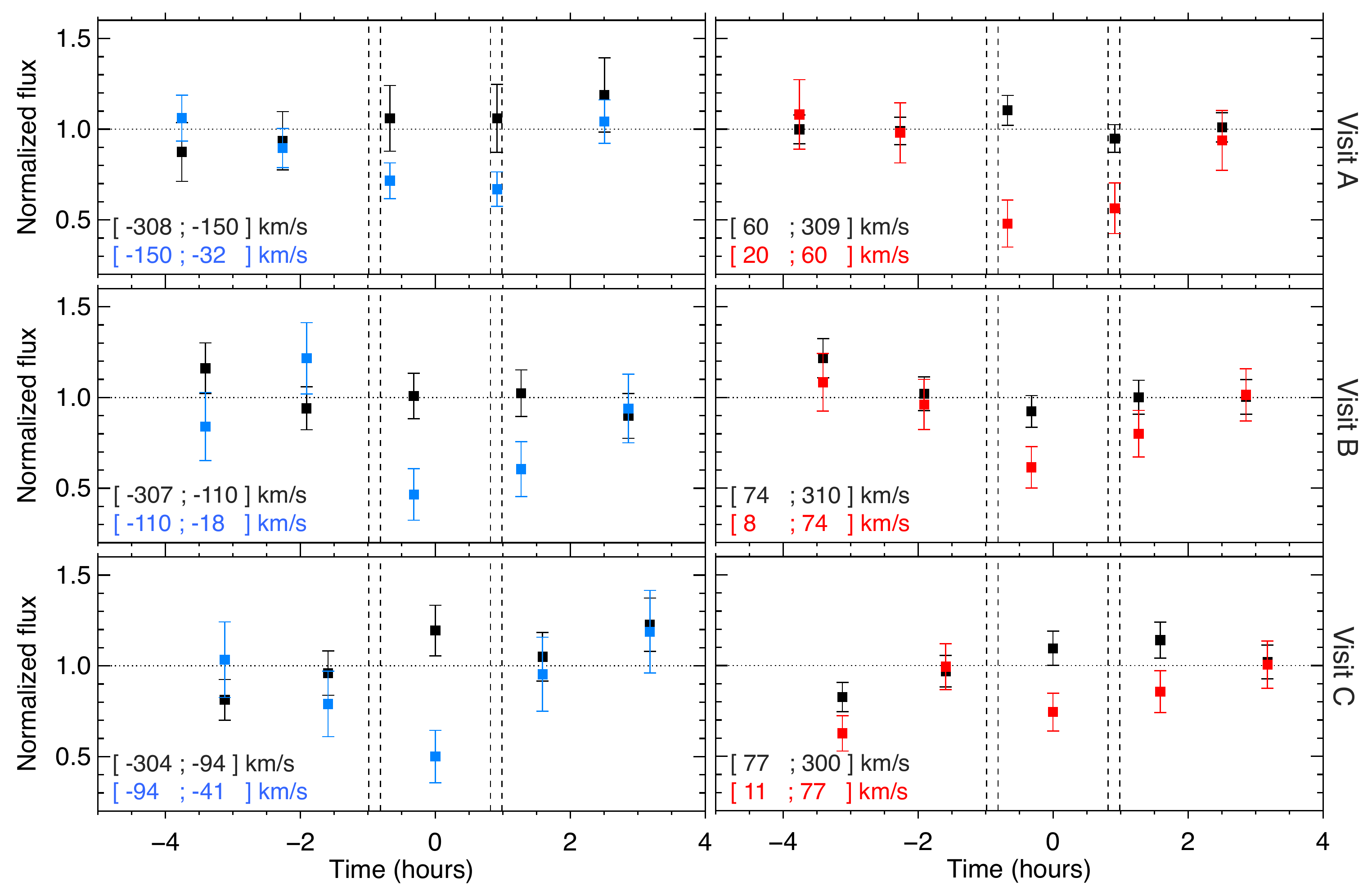}
\end{minipage}
\caption[]{Lyman-$\alpha$ line flux integrated over bands of interest in visits A, B, and C, as a function of time relative to the mid-transit of GJ\,3470b (contacts are shown as vertical dashed lines). The spectral ranges of the bands are indicated in each sub-panel. For all visits, blue and red points correspond to bands showing absorption in the blue and the red wing, respectively. We note that the bands were optimized for all orbits in a visit, and do not necessarily show the strongest absorption in a given orbit. Black points correspond to the complementary of the absorbed regions up to the limit of the wings. Visits A and B show stable flux levels in the wings, except for a flux spike at the peak of the red wing in Visit B orbit 1, while Visit C displays short-term variability. Fluxes are normalized by the average flux over orbits 1-2-5 (excluding spurious ranges in the red wing for visits B and C, see text).} 
\label{fig:LC_grid}
\end{figure*}

\begin{figure}     
\includegraphics[trim=0cm 0cm 0cm 0cm,clip=true,width=\columnwidth]{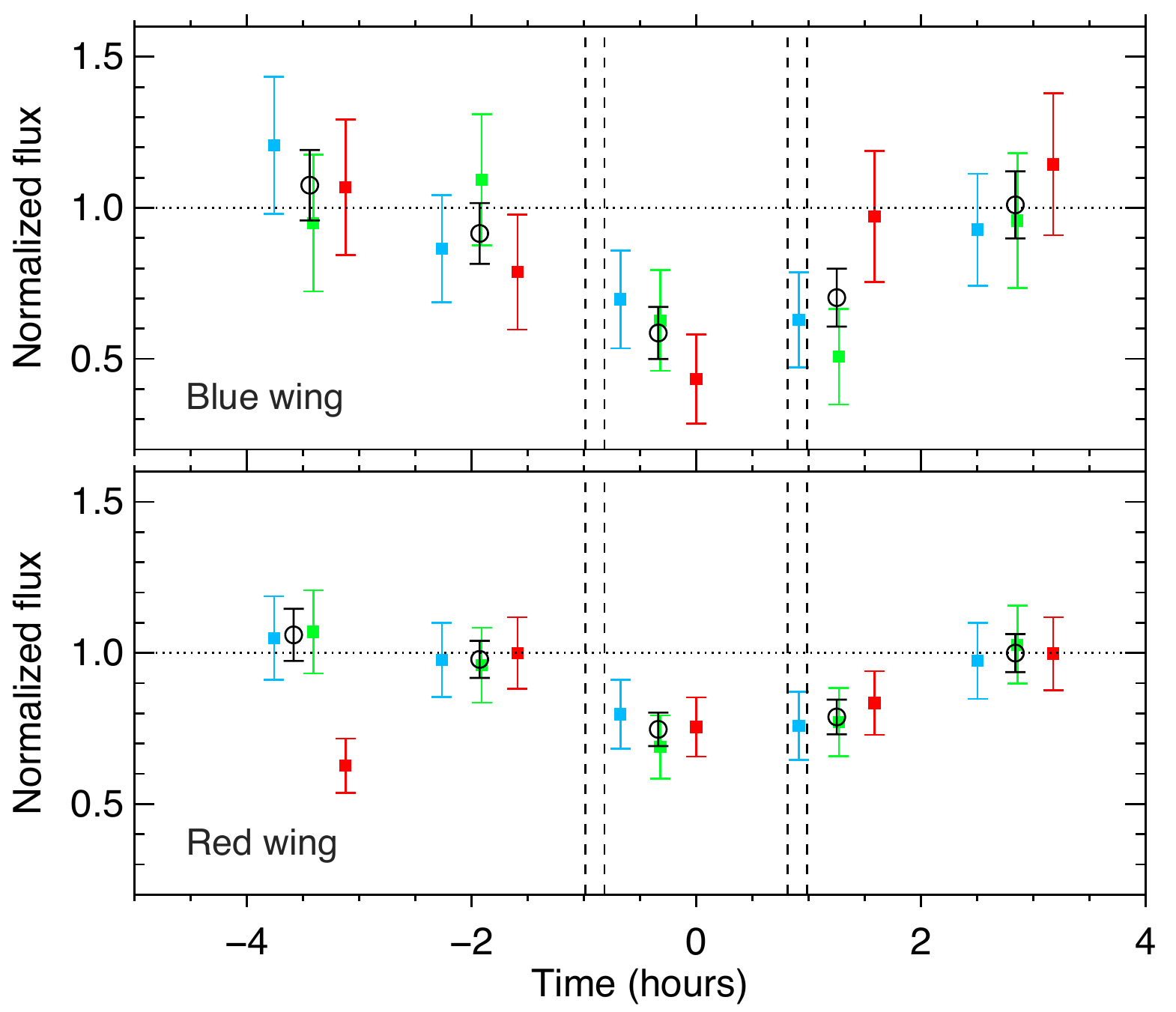}
\caption[]{Flux integrated in the blue wing of the Lyman-$\alpha$ line over -94 to -41\,km\,s$^{-1}$ (upper panel), and in the red wing over 23 to 76\,km\,s$^{-1}$ (lower panel). Visits A, B, and C are plotted in blue, green, and red. Fluxes are normalized by the average flux over orbits 1, 2, and 5. Time is relative to the mid-transit of GJ\,3470b (contacts are shown as vertical dashed lines). Exposures occuring at similar times in each visit have been binned into the black points. Orbit 1 in Visit C has been excluded from the normalisation and the binning because of its spurious flux decrease between +38 and +116\,km\,s$^{-1}$.}
\label{fig:LC_allVis}
\end{figure}

The stellar Lyman-$\alpha$ line profile of GJ\,3470 remains mostly stable within each visit (but see Sect.~\ref{sec:Ly_rec} across the visits). Absorption signatures are detected in both wings of the Lyman-$\alpha$ line in all visits. They are located in similar spectral ranges between about -150 and +90\,km\,s$^{-1}$ at maximum, although absorption beyond -100\,km\,s$^{-1}$ is only measured in Visit A. The core of GJ\,3470 Lyman-$\alpha$ line cannot be analyzed because it is fully absorbed by the interstellar medium (ISM). Nonetheless, we measure absorption down to the airglow boundaries (varying between -40 and +20\,km\,s$^{-1}$ due to the differences in airglow position and width between visits), which suggests that it extends to lower absolute velocities. This would be expected if the planetary atmosphere is the source for the absorbing neutral hydrogen atoms, a large fraction of which would move with low radial velocities near the transit when they are still in the vicinity of the planet. The light curves plotted over the absorbed spectral ranges in Fig.~\ref{fig:LC_allVis} show that the Lyman-$\alpha$ line transit keeps a similar shape over the three visits (absorption depths are reported in Table~\ref{tab:sum_abs}). Apart from spurious variations in the first orbits of Visits B and C, likely stellar in origin, spectra obtained before -1.5\,h and after 2\,h show no variations in any of the visits, indicating the absence of neutral hydrogen occulting the star long before or after the transit. In contrast, absorption is consistently detected during and shortly after the planetary transit (Fig.~\ref{fig:LC_grid}).\\

\begin{table}[tbh]
\caption{Absorption depths measured in the blue (-94 to -41\,km\,s$^{-1}$) and red wings (23 to 76\,km\,s$^{-1}$) of the Lyman-$\alpha$ line.}
\centering
\begin{tabular}{lcccc}
\hline
\hline
\noalign{\smallskip}
					& \multicolumn{2}{c}{Blue wing} 	& \multicolumn{2}{c}{Red wing} \\	
					&	 In-transit & Egress	  &	 In-transit & Egress	  \\	
\noalign{\smallskip}
\hline
\noalign{\smallskip}
Visit A & 30.7$\pm$16.3    & 36.5$\pm$15.9  & 19.0$\pm$11.2 &	24.4$\pm$10.9	 \\
Visit B & 37.8$\pm$16.7     & 49.1$\pm$16.0 & 29.5$\pm$10.3 &	23.6$\pm$10.9	 \\
Visit C & 56.9$\pm$14.8     & 2.5$\pm$22.0  & 24.0$\pm$9.6 &	17.2$\pm$10.2	 \\
\noalign{\smallskip}
\hline
\noalign{\smallskip}
All visits & 41.8$\pm$8.7     & 29.4$\pm$9.7  & 24.2$\pm$5.4 &	21.8$\pm$5.5	 \\
\noalign{\smallskip}
\hline
\hline
\end{tabular}
\label{tab:sum_abs}
\end{table}

\begin{figure}     
\includegraphics[trim=2cm 2.cm 1cm 6cm,clip=true,width=\columnwidth]{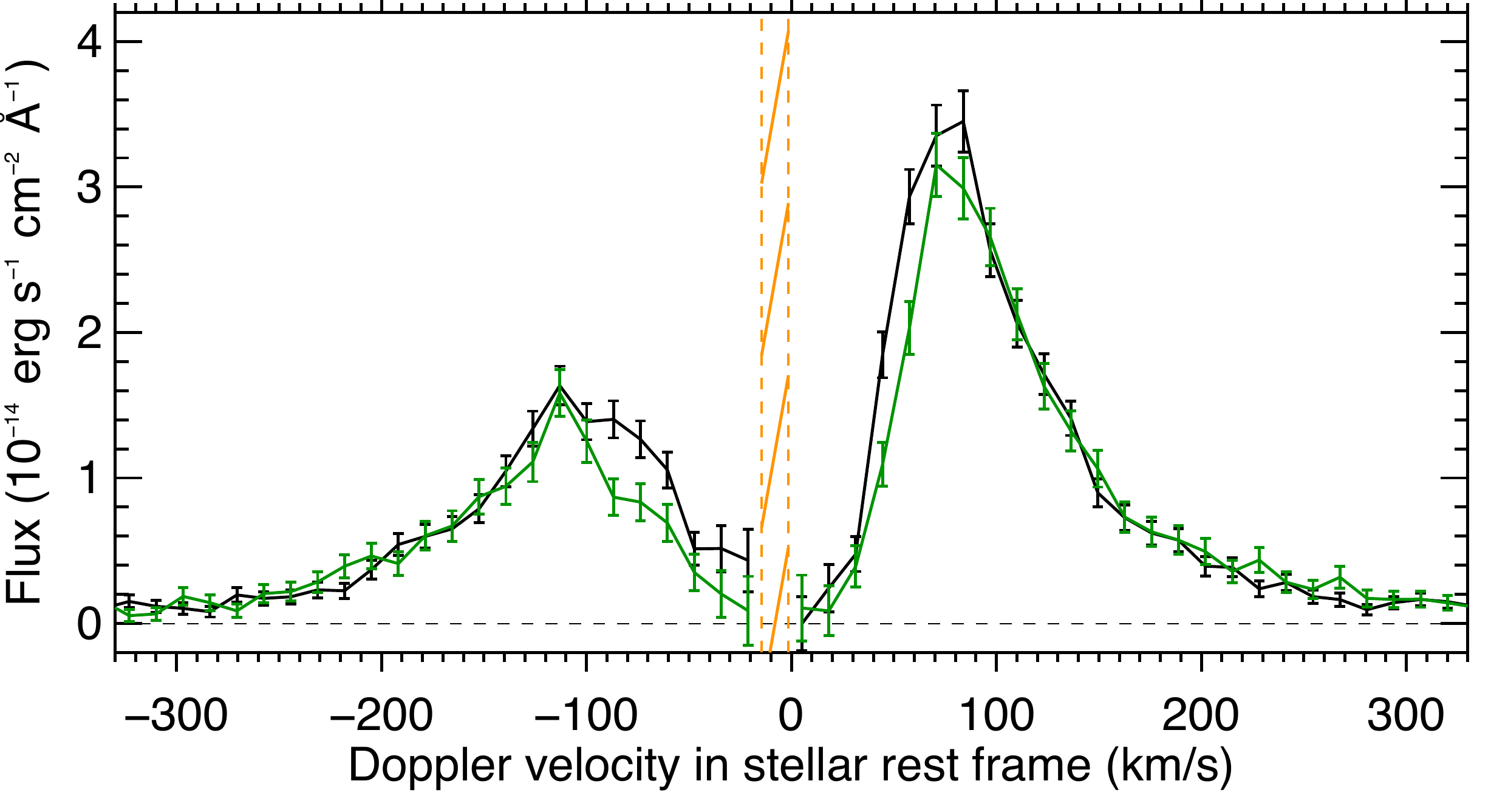}
\caption[]{Average Lyman-$\alpha$ line spectra of GJ\,3470b over the three visits, during the transit of the extended neutral hydrogen atmosphere (green spectrum) and outside (black spectrum). The dashed orange region is contaminated by the airglow in  all visits.}
\label{fig:Spec_av_allVis}
\end{figure}

The repeatability of the absorption signatures, their spectral ranges spread around the planet radial velocity ($\sim$5\,km\,s$^{-1}$ at mid-transit), and their phasing with the planetary transit, allow us to conclude the detection of an extended upper atmosphere of neutral hydrogen around the warm-Neptune GJ\,3470 b. We plot in Fig.~\ref{fig:Spec_av_allVis} the absorbed and stable spectra averaged over the three visits to highlight the signature from this extended atmosphere. We measure absorption depths of 35$\pm$7\% in the blue wing, between -94 and -41\,km\,s$^{-1}$, and 23$\pm$5\% in the red wing, between 23 and 76\,km\,s$^{-1}$ (Table~\ref{tab:sum_abs}). \\

\section{The Lyman-$\alpha$ line of GJ\,3470}
\label{sec:Ly_rec}

\subsection{Reconstruction of the intrinsic line profiles}

The profile of the intrinsic stellar Lyman-$\alpha$ line, before absorption by the ISM, is required to calculate the radiation pressure acting on escaping neutral hydrogen atoms, and can be used to estimate the stellar UV emission (see Sect.~\ref{sec:em_lines}). We used the out-of-transit spectra taken as reference to reconstruct the intrinsic stellar Lyman-$\alpha$ line in each visit. To use as much information as possible, we included the stable wings of the in-transit spectra where no absorption was detected. We follow the same procedure as in previous studies (e.g. \citealt{Bourrier2015_GJ436,Bourrier2017_HD976}): a model profile of the intrinsic stellar line is absorbed by hydrogen and deuterium in the ISM, convolved with STIS line spread function (LSF), and fitted to the data. The model is oversampled in wavelength, and rebinned over the STIS spectral table after convolution. \\

\begin{figure}[tbh!]
\centering
\includegraphics[trim=0cm 0cm 0cm 0cm,clip=true,width=\columnwidth]{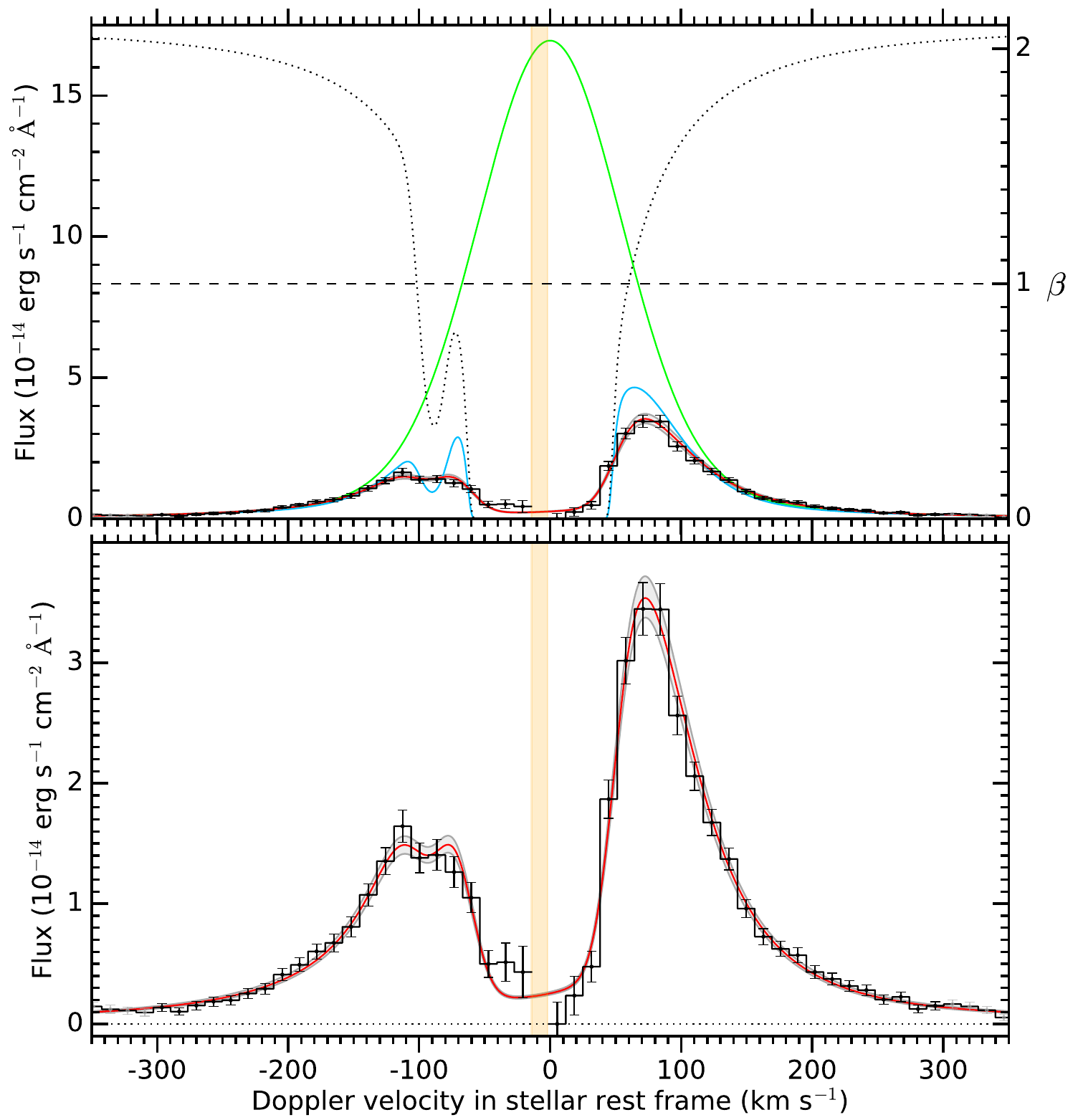}
\caption[]{Lyman-$\alpha$ line profile of GJ\,3470 averaged over the three visits. The black histogram shows the observed spectra, fitted over points with black error bars. The orange band shows the regions too contaminated by geocoronal emission to be included in the fit. The green line is the best-fit for the intrinsic stellar line profile. It yields the blue profile after absorption by the interstellar medium, whose profile is plotted as a dotted black line (ISM absorption in the range 0-1 has been scaled to the vertical axis range). The red line shows the line profile fitted to the data, after convolution with the STIS LSF. Grey bands delimitate the 1$\sigma$ envelopes of the best-fit profiles. The theoretical intrinsic stellar line profile also corresponds to the profile of the ratio $\beta$ between radiation pressure and stellar gravity, which is reported on the right axis in the top panel. The bottom panel shows a zoom of the observed Lyman-$\alpha$ line and its best-fit model.}
\label{fig:La_stellar_lines}
\end{figure}

We found that a single-peaked Voigt profile provides the best fit to the intrinsic Lyman-$\alpha$ line, compared to other typical line profiles with two peaks or Gaussian components (e.g. \citealt{Wood2005}). This strengthens the trend noted in \citet{Bourrier2017_K444} that M dwarfs show no self-reversal in their core, while K dwarfs sometimes show double-peaked profiles and G-type stars always do. K dwarfs might thus have transition regions, where the core of the Lyman-$\alpha$ line is formed (\citealt{Vernazza1981}), that have intermediate structures in between those of G-type stars and M dwarfs. We fitted the spectra from all visits together using the Markov-Chain Monte Carlo (MCMC) Python software package \textit{emcee} (\citealt{Foreman2013}). The theoretical intrinsic line profile is specific to each visit and is defined by its centroid $\gamma_{\mathrm{Ly-\alpha}}$, its temperature $T_{\mathrm{Ly-\alpha}}$ (assuming pure thermal broadening), its damping parameter $a_\mathrm{damp}$, and its total flux $F_{\mathrm{Ly-\alpha}}$(1\,au). The absorption profile of the ISM along the line-of-sight is common to all visits, and defined by its column density of neutral hydrogen log$_{10}$\,$N_{\mathrm{ISM}}$(H\,{\sc i}), its Doppler broadening parameter $b_{\mathrm{ISM}}$(H\,{\sc i}), and its heliocentric radial velocity $\gamma_{\mathrm{ISM}/\sun}$. The D\,{\sc i}/H\,{\sc i} ratio was set to 1.5$\times$10$^{-5}$\ ({e.g.} \citealt{Hebrard_Moos2003}; \citealt{Linsky2006}). Best-fit values for the model parameters are given in Table~\ref{table:tab_paramsfit}, yielding a good $\chi^2$ of 95 for 113 degrees of freedom (128 datapoints and 15 free parameters; Fig.~\ref{fig:La_stellar_lines}).\\

\subsection{Results and interpretation}
\label{sec:Ly_line_results}

The LISM Kinematic Calculator\footnote{\mbox{\url{http://sredfield.web.wesleyan.edu/}}}, a dynamical model of the local ISM (\citealt{Redfield_Linsky2008}) predicts that the line-of-sight (LOS) toward GJ\,3470 crosses the LIC and Gem cloud with heliocentric radial velocities of 18.3\,km\,s$^{-1}$ and 33.4\,km\,s$^{-1}$, respectively. The velocity of the ISM cloud we fitted is in good agreement with that of the Local Interstellar Cloud (LIC), and its column density (log$_{10}$\,$N_{\mathrm{ISM}}$(H\,{\sc i}) = 18.07$\pm$0.09\,cm$^{-2}$) is consistent with the range of values expected for a star at a distance of 29.5\,pc (Fig. 14 in \citealt{Wood2005}). \\

The total flux in the observed and reconstructed Lyman-$\alpha$ lines show no significant variations between the three visits (Table~\ref{table:tab_paramsfit}). The best-fit model for the line averaged over the three visits is shown in Fig.~\ref{fig:La_stellar_lines}. It yields a total flux F$_{\mathrm{Ly-\alpha}}$(1\,au) = 3.64$\pm$0.15\,erg\,cm$^{-2}$\,s$^{-1}$, which is in remarkable agreement with the value of 3.7\,erg\,cm$^{-2}$\,s$^{-1}$ derived from the empirical relation in \citet{Youngblood2016} and the rotation period of GJ\,3470 (Table~\ref{tab:param_sys}). We note that uncertainties in the Lyman-$\alpha$ line reconstruction do not affect the measurement of GJ\,3470b absorption signature, which is a relative measure between in- and out-of-transit spectra both absorbed in the same way by the interstellar medium. \\

Our reconstruction for the intrinsic Lyman-$\alpha$ line of GJ\,3470 shows that radiation pressure is strong enough for neutral hydrogen atoms escaping from the planet to be accelerated away from the star. Indeed, radiation pressure from this M1.5 dwarf overcomes its gravity by up to a factor 2 (Fig.~\ref{fig:La_stellar_lines}), which is intermediate between the factor 3 associated with the K0 dwarfs HD\,189733 and HD\,97658 (\citealt{Bourrier_lecav2013}; \citealt{Bourrier2017_HD976}) and the factor 0.7 associated with the M2.5 dwarf GJ\,436 (\citealt{Bourrier2015_GJ436}). We thus expect the putative hydrogen exosphere of GJ\,3470 b to be in an intermediate regime of radiative blow-out, in between the stronger radiative blow-out acting on the comet-like exospheres of HD\,209458 (\citealt{Lecav2008}) or HD\,189733b (\citealt{Bourrier_lecav2013}), and the radiative braking shaping the giant exosphere of GJ\,436b (\citealt{Bourrier2015_GJ436, Bourrier2016}).  \\


\section{XUV irradiation of the upper atmosphere}
\label{sec:em_lines}

\begin{figure*}
\centering
\begin{minipage}[h!]{\textwidth}
\includegraphics[trim=0cm 0cm 0cm 0cm,clip=true,width=\columnwidth]{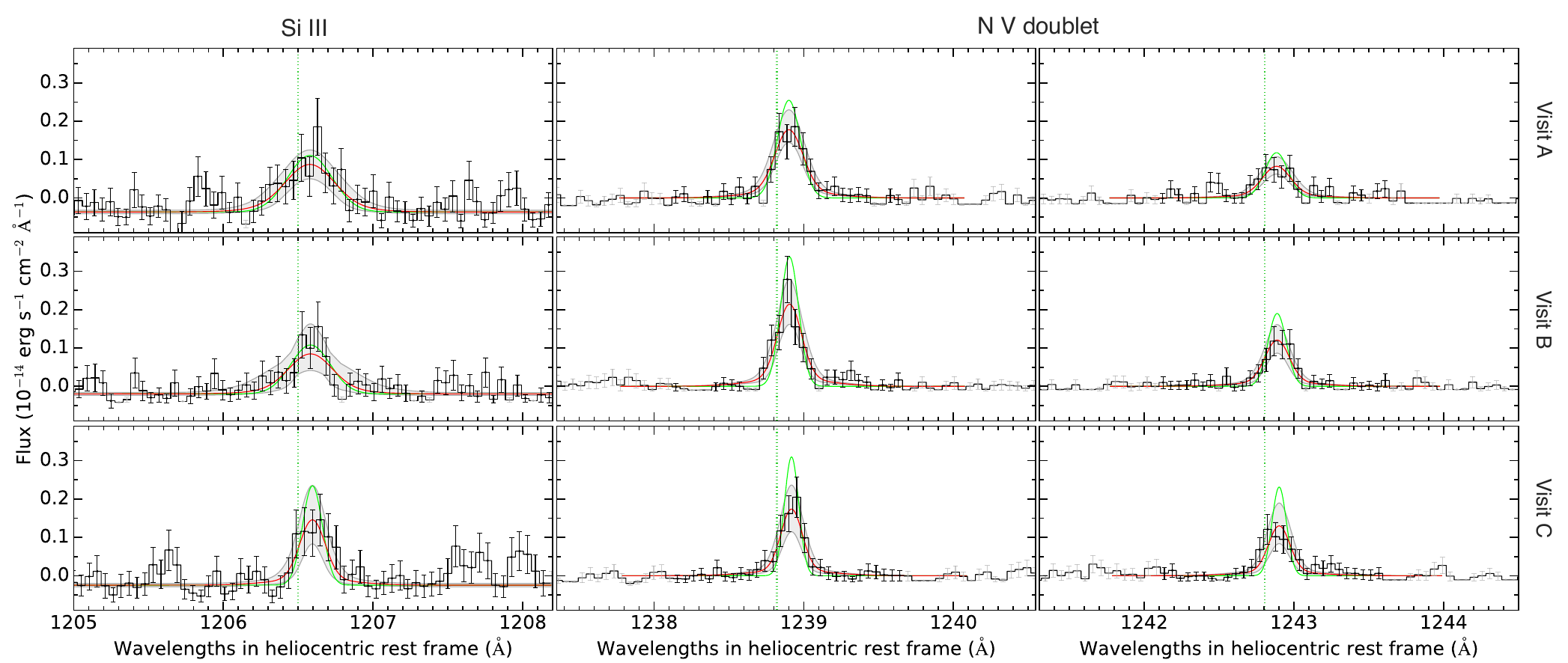}
\end{minipage}
\caption[]{Spectral profiles of GJ\,3470 \ion{Si}{iii} line and \ion{N}{v} doublet. Each row corresponds to a visit. Black histograms show the observed spectra, fitted over points with black error bars. The green lines are the best fits for the intrinsic stellar line profile, which yield the red line after convolution with the STIS LSF. Grey bands delimitate the 1$\sigma$ envelopes of the best-fit profiles. Vertical dashed lines identify the lines transitions in the Sun rest frame.}
\label{fig:stellar_lines}
\end{figure*}

Besides the Lyman-$\alpha$ line (1215.67\,\AA), we identified the stellar emission lines of \ion{Si}{iii} (1206.5\,\AA, $\log\,T_\mathrm{max}\rm (K)$ = 4.8) and the \ion{N}{v} doublet (1242.8\,\AA\, and 1238.8\,\AA, $\log\,T_\mathrm{max}\rm (K)$ = 5.3) in the STIS spectra of GJ\,3470. These lines are useful tracers of the stellar transition region. To measure their strengths, we averaged the spectra in each visit and fitted Gaussian models to the three emission lines following the same approach as in Sect.~\ref{sec:Ly_rec}. We assumed that there is no turbulent broadening and that the lines share a common Doppler shift. We allowed for a different continuum level in the range of the \ion{Si}{iii} line because we found it was overcorrected by CALSTIS to negative values. The best-fit properties for the lines are reported in Table~\ref{table:tab_paramsfit} and the corresponding models are displayed in Fig.~\ref{fig:stellar_lines}. The radial velocity of the Lyman-$\alpha$, \ion{Si}{iii}, and \ion{N}{v} lines are consistent in each visit, but they are systematically blueshifted with respect to the heliocentric radial velocity of GJ\,3470. This is not the expected behaviour for high-energy stellar emission lines, which can possibly show physical redshifts due to the structure of the stellar chromosphere (e.g. \citealt{Linsky2012}). It is therefore likely that these marginal shifts arise from an offset in STIS wavelength calibration. We accounted for this offset in the fit to the ISM properties and in the interpretation of the observed spectra hereafter.\\

\begin{figure}
\centering
\includegraphics[trim=0cm 0cm 0cm 0cm,clip=true,width=\columnwidth]{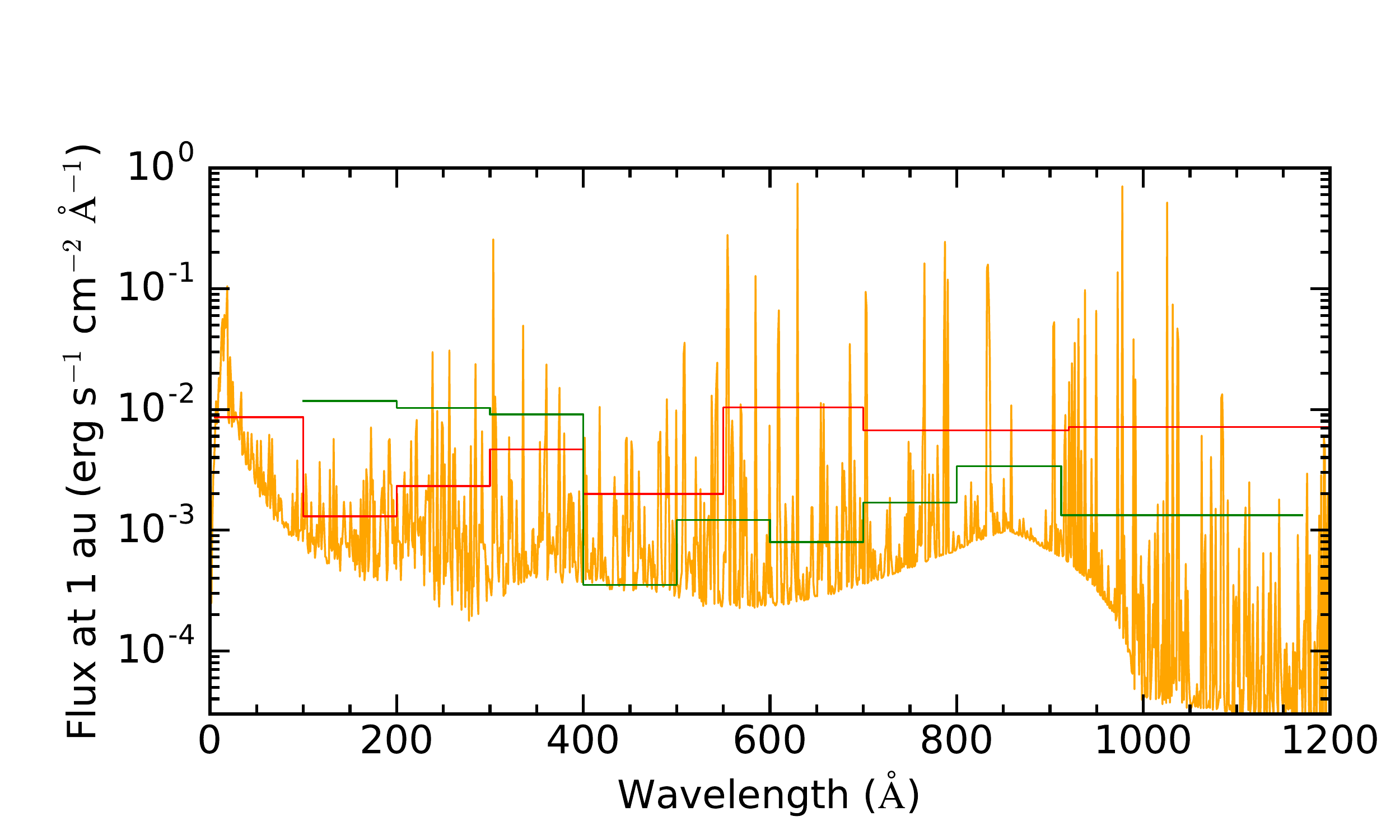}
\caption[]{Synthetic XUV spectrum of GJ\,3470 (orange) derived from the coronal model, as explained in the text. The spectrum is integrated over broad spectral bands (red) for comparison with the values derived from the semi-empirical relations in \citet{Linsky2014} using our measured Lyman-$\alpha$ flux (green).}
\label{fig:XUV_spec}
\end{figure}

Most of the stellar EUV spectrum is not observable from Earth because of ISM absorption. We have thus constructed a synthetic XUV spectrum of GJ\,3470 in the region 1-1200\,\AA\, based on a coronal model, following \citet{Sanz-Forcada2011}. We base our coronal model on the X-ray spectra obtained from the XMM-Newton archive (obs ID 763460201, observed in 2015/04/15, P.I. Salz) for the coronal region, and on the UV fluxes derived for the \ion{Si}{iii} and \ion{N}{v} lines for the transition region (we used their weighted mean since no significant variations were detected between the three visits). The spectral fit gives an X-ray (0.12-2.48\,keV, or
5-100\,\AA) luminosity of 2.3$\times$10$^{27}$\,erg\,s$^{-1}$. The model-derived luminosities in different EUV and FUV bands are 1.2$\times$10$^{28}$\,erg\,s$^{-1}$ (100-920\,\AA), 5.6$\times$10$^{27}$\,erg\,s$^{-1}$ (920-1200\,\AA) and 2.7$\times$10$^{27}$\,erg\,s$^{-1}$ (100-504\,\AA). The coronal model also predicts a Lyman-$\alpha$ flux of 7.0\,erg\,cm$^{-2}$\,s$^{-1}$ at 1\,au, \textit{i.e.} a factor of $\sim$2 only of difference with the flux derived from the observations (Table~\ref{table:tab_paramsfit}). This is remarkable, considering that the coronal model is not expected to correctly predict the Lyman-$\alpha$ flux. Further details about the coronal model will be given in Sanz-Forcada et al. (in prep.). We show in Fig.~\ref{fig:XUV_spec} the comparison between our XUV spectrum and the broadband fluxes obtained with the semi-empirical relations from \citet{Linsky2014} and the intrinsic Lyman-$\alpha$ fluxes derived from our observations. Lyman-$\alpha$ is formed in a limited range of temperature in the transition region, thus the use of a coronal model based on actual data should always be preferred to correctly sample the transition region and coronal temperatures where emission lines in the XUV band are
originated (\citealt{Sanz-Forcada2015}).  \\

The XUV spectrum derived between 5 and 911.8\,\AA\, yields a photoionization lifetime of 55\,min for neutral hydrogen atoms at the semi-major axis of GJ\,3470 b (details on the calculation can be found in \citealt{Bourrier2017_HD976}). For comparison, we obtain a lifetime of 2.6\,h when using the EUV flux derived from the  \citet{Linsky2014} relations. While other processes like radiation pressure or the strength of the planetary outflow play a role in shaping exospheres, the short photoionization lifetime at GJ\,3470b likely prevents escaping hydrogen atoms from remaining neutral as far away from the planet as in the giant exosphere of GJ\,436 b (photoionization lifetime $\sim$12\,h).

\begin{table*}[tbh]
\caption{Properties derived from the fit to GJ\,3470 stellar lines.}
\centering
\begin{threeparttable}
\begin{tabular}{lllccll}
\hline
\noalign{\smallskip}  
 &    &   \textbf{Parameter}       & \textbf{Visit 1}      & \textbf{Visit 2}         & \textbf{Visit 3} 		   & \textbf{Unit} \\   			
\noalign{\smallskip}
\hline
\hline
\noalign{\smallskip}
\textbf{Stellar lines} & Radial velocity &  $\gamma_{\ion{Si}{iii}+\ion{N}{v}}$    &  19.2$\pm$3.4      & 20.4$\pm$2.6              &  23.5$\pm$2.7	& km\,s$^{-1}$  \\
      &       &    $\gamma_{\mathrm{Ly-\alpha}}$    &  21.8$\pm$1.7      & 22.4$\pm$1.6  &  25.0$\pm$1.8	& km\,s$^{-1}$  \\   
 & Total flux & F$_{\ion{Si}{iii}}$(1\,au)   & 	20.2$\pm$4.1   & 15.9$\stackrel{+4.0}{_{-3.2}}$		   & 15.2$\pm$3.5  & 10$^{-3}$\,erg\,cm$^{-2}$\,s$^{-1}$\\                 
&			& F$_{\ion{N}{v}}^{1238.8}$(1\,au)   & 18.0$\pm$2.4		  & 19.0$\stackrel{+2.4}{_{-2.8}}$ 		   & 13.6$\pm$2.1 & 10$^{-3}$\,erg\,cm$^{-2}$\,s$^{-1}$\\                 
&		& F$_{\ion{N}{v}}^{1242.8}$(1\,au)   & 8.3$\pm$1.7		  & 10.7$\pm$1.8 		   & 10.2$\stackrel{+1.7}{_{-1.8}}$  & 10$^{-3}$\,erg\,cm$^{-2}$\,s$^{-1}$\\                 
&		& F$_{\mathrm{Ly-\alpha}}$(1\,au)   & 3.4$\stackrel{+0.7}{_{-0.5}}$		  & 3.5$\stackrel{+0.7}{_{-0.5}}$   & 4.2$\stackrel{+0.9}{_{-0.7}}$  & \,erg\,cm$^{-2}$\,s$^{-1}$\\                   
 & Temperature & T$_{\ion{Si}{iii}}$   & 47$\stackrel{+33}{_{-21}}$   & 35$\stackrel{+106}{_{-28}}$ 	   & 9.0$\stackrel{+6.9}{_{-5.1}}$   & 10$^{5}$\,K\\                 
 & 			& T$_{\ion{N}{v}}$   & 5.8$\stackrel{+2.6}{_{-2.1}}$  		  & 3.7$\stackrel{+2.2}{_{-1.6}}$		   & 2.3$\pm$1.6  & 10$^{5}$\,K\\
 & 			& T$_{\mathrm{Ly-\alpha}}$   & 3.9$\stackrel{+0.6}{_{-0.5}}$  		  & 3.1$\pm$0.4		   & 2.7$\stackrel{+0.4}{_{-0.3}}$  & 10$^{5}$\,K\\ 
\noalign{\smallskip} 
\hline
\noalign{\smallskip} 
\textbf{ISM} & Radial velocity & $\gamma_{\mathrm{ISM}/\sun}$ & \multicolumn{3}{c}{18.8$\stackrel{+1.4}{_{-1.5}}$}  & km\,s$^{-1}$  \\ 
			 & Column density  & log$_{10}$\,$N_{\mathrm{ISM}}$(H\,{\sc i})  & \multicolumn{3}{c}{18.07$\pm$0.09}  & cm$^{-2}$   \\ 
 			 & Doppler broadening & $b_{\mathrm{ISM}}$(H\,{\sc i}) & \multicolumn{3}{c}{16.6$\stackrel{+0.6}{_{-0.7}}$}  & km\,s$^{-1}$  \\  		                 		
\hline
\noalign{\smallskip}
\hline
\end{tabular}
  \end{threeparttable}
\label{table:tab_paramsfit}
\end{table*}


\section{3D simulations of GJ\,3470 b upper atmosphere}

\subsection{Modeling with the EVE code}
\label{sec:descr_EVE}

We interpreted the Lyman-$\alpha$ transit spectra of GJ\,3470 b using 3D simulations of its upper atmosphere with the EVaporating Exoplanets (EVE) code (\citealt{Bourrier_lecav2013, Bourrier2014, Bourrier2015_GJ436, Bourrier2016}). In this code, a planetary system is simulated over time in the star rest frame. A parameterized grid, corresponding to the collisional layers of the atmosphere, is described with a 3D grid in which the density, velocity, and temperature of neutral hydrogen are calculated using analytical functions. The upper boundary of the parameterized atmosphere represents the altitude beyond which species decouple in the collision-less exosphere. Metaparticles of neutral hydrogen, representing groups of atoms with the same properties, are released at this boundary. Their dynamics is calculated using a particle Monte-Carlo simulation that accounts for the planet and star gravity, the inertial force linked to the non-Galilean stellar reference frame, radiation pressure, and interactions with the stellar wind (see details in the above papers). The population of neutral hydrogen atoms in the exosphere follows an exponential decay because of stellar photoionization, which is calculated using an input XUV spectrum. Particles in the shadow of the opaque planetary disk are not interacting with photons and particles from the star. \\

At each timestep of the simulation the code calculates the theoretical spectrum seen from Earth as the sum of the local spectra over the discretized stellar disk, accounting for the opaque planetary disk occultation and for the wavelength-dependent absorption from the bottom atmosphere, the exosphere, and the ISM. Atmospheric cells and metaparticles in front of, or behind, the opaque planetary disk do not contribute to the absorption. Theoretical spectra are calculated with a high spectral resolution and temporal cadence compared to the observations. At the end of the simulation, they are downsampled to the resolution of the observed spectra and averaged within the time window of the observed exposures, weighting their relative contribution by the effective overlap with the observed time window. The final spectra, convolved with the instrumental response, are comparable with the observations. Modeling with the EVE code thus allow us to use all information available in an observed dataset, accounting for the partial occultation of the stellar disk during ingress/egress and for 3D effects inherent in the asymetrical structure of extended atmospheres.


\subsection{Application to GJ\,3470b}

There are two remarkable features in the observations, which we detail below.

(i) Absorption is observed up to high positive velocities ($\sim$90\,km\,s$^{-1}$) during and shortly after the optical transit. This is challenging to explain by neutral hydrogen gas infalling toward the star. Indeed, upper atmospheric layers sheared by the stellar gravity would stream through the L1 Lagrangian point ahead of the planet and occult the star before its transit, not after (e.g. \citealt{Bourrier2015_GJ436}). Interactions between the stellar and planetary magnetic fields could possibly force ionized planetary material to stream toward the star behind the planet, carrying along neutral hydrogen (\citealt{Matsakos2015,Strugarek2016}), but radiation pressure or interactions with the stellar wind might stop the infalling flow of neutral hydrogen (e.g. \citealt{Bisikalo2013}). 

Alternatively, the Lorentzian wings from the absorption profile of a dense, extended thermosphere (or damping wings) have been proposed as a possible origin for redshifted Lyman-$\alpha$ absorption signatures from exoplanets (\citealt{Tian2005}; \citealt{BJ2008}). For damping wings to explain the observations of GJ\,3470b, its thermosphere would however have to show a larger extension along the planetary motion than in the perpendicular directions. Indeed, a spherical thermosphere large enough to yield absorption up to 1.5\,h after the transit would yield a much deeper absorption than observed at mid-transit (see Fig.~\ref{fig:LC_allVis} and next section). To test this scenario, we modified the EVE code so that it can simulate the thermosphere as an ellipsoid with radius $R_{||}$ along the tangent to its orbit, $R_{\perp}^\mathrm{orb}$ along the normal to this tangent in the orbital plane, and $R_{\perp}^\mathrm{norm}$ along the perpendicular to the orbital plane. The ellipsoid can be off-centered from the planet along the tangent to its orbit by $d_{||}$ (positive when the ellipsoid center is ahead of the planet). Because we are only probing column densities along the LOS, we did not try to fit possible inclination of the ellipsoidal atmosphere with respect to the orbital trajectory. We assumed a simple isotropic, hydrostatic density profile, defined by a mean temperature $T_\mathrm{th}$ (in K) and the density of neutral hydrogen $n_\mathrm{th}$ at 1\,$R_\mathrm{p}$ (in atoms/cm$^{3}$). We assumed a Solar-like composition of atomic hydrogen and helium ($\mu$ = 1.2).\\
	 
(ii) Absorption is deeper in the blue wing and extends to higher absolute velocities than in the red wing. This spectral asymetry is even more pronounced in the planet rest frame, since the radial orbital velocity of GJ\,3470~b is positive during the transit. Blueshifted Lyman-$\alpha$ absorption signatures are a telltale sign of neutral hydrogen exospheres around evaporating planets (e.g. \citealt{VM2003}; \citealt{Lecav2012}; \citealt{Ehrenreich2015}), and we thus modeled the exosphere of GJ\,3470b using particle simulations of the escaping gas (Sect.~\ref{sec:descr_EVE}). Metaparticles of neutral hydrogen are released uniformely from the surface of the ellipsoidal atmosphere, with an upward velocity of 5\,km\,s$^{-1}$ typical of outflows from low-gravity planets (\citealt{Salz2016b}). Because the orbit of GJ\,3470b is eccentric, the mass loss rate of neutral hydrogen was set to vary as the inverse square of the distance to the star, with the fitted value $\dot{M}_{H^{0}}$ (in g\,s$^{-1}$) defined at the semi-major axis. The evolution of neutral hydrogen atoms is calculated under radiation pressure and photoionization using the Lyman-$\alpha$ and XUV spectra derived in Sect.~\ref{sec:Ly_rec} and ~\ref{sec:em_lines}. \\

The observed and simulated time-series spectra were compared over the velocity range -300 to 300\,km\,s$^{-1}$ (in the star rest frame), excluding the airglow-contamined regions and the spurious regions in the first exposures of visits B and C (Sect.~\ref{sec:sum_obs}). As shown in the following sections, the detected signature is well explained by radiation pressure acceleration with no evidence for interactions with the stellar wind. We thus followed the same approach as in previous studies (e.g. \citealt{Bourrier_lecav2013,Bourrier2014,Bourrier2016}), which is to fit unknown mechanisms in our simulations only when the known forces are not sufficient to explain the observations. In contrast to radiation pressure, derived directly from the observed Lyman-$\alpha$ line, there are only model-derived estimates for the wind properties of exoplanet host stars and M dwarfs winds in particular are poorly known (\citealt{Vidotto2017}). In summary, the free parameters in the model are the 3D dimensions of the ellipsoidal atmosphere $R_{||}$, $R_{\perp}^\mathrm{orb}$, $R_{\perp}^\mathrm{norm}$ and its offset with respect to the planet center $d_{||}$, the temperature $T_\mathrm{th}$ and density $n_\mathrm{th}$ (at 1\,$R_\mathrm{p}$) of neutral hydrogen in the upper atmosphere, and its mass loss rate $\dot{M}_{H^{0}}$ at the exobase. We explored a grid of these parameters, using the $\chi^{2}$ as the merit function to identify a best-fit model and study the influence of the atmospheric properties.  \\


\subsection{The structure of GJ\,3470 b extended atmosphere}
\label{sec:atm_struct}

\begin{figure*}
\begin{minipage}[h!]{\textwidth}
\centering
\includegraphics[trim=0cm 0cm 0cm 0cm,clip=true,width=0.7\columnwidth]{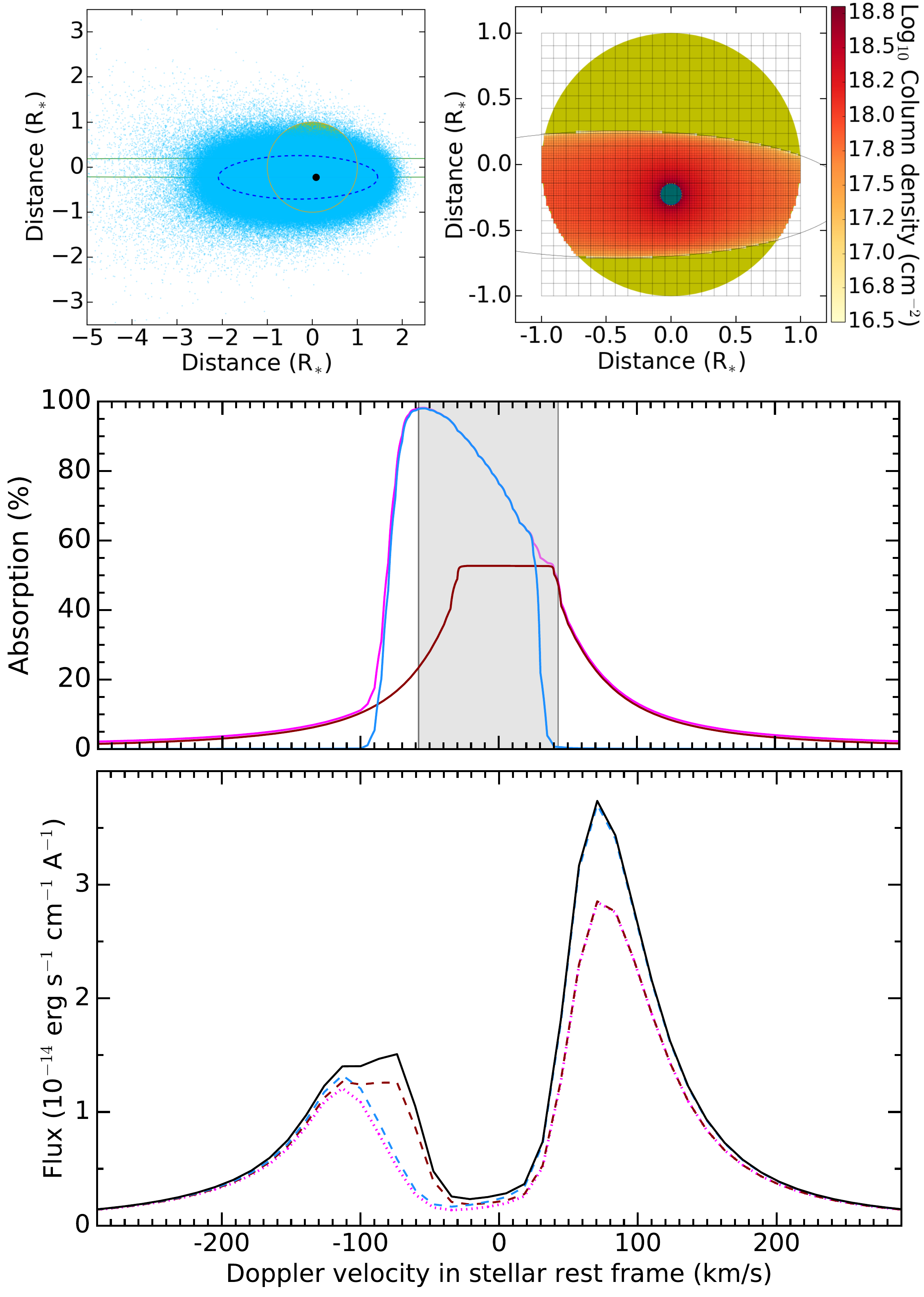}
\end{minipage}
\caption[]{Contributions to the absorption by the different regions of GJ\,3470b upper atmosphere at mid-transit, from one of the best-fit simulations. \textit{Top panel :} Transit view of the exosphere (left) and of the ellipsoidal atmosphere (right). In the left subpanel are shown in projection in the plane of sky the planetary orbit (green curve) and the boundary of the ellipsoidal atmosphere (dashed blue curve). The planet as seen at optical wavelengths is shown as a black and a blue disk in the left and right subpanels, respectively. \textit{Middle panel :} Total absorption profile from GJ\,3470b (magenta), and independent contributions from its exosphere (blue) and ellipsoidal atmosphere (dark red), with neither ISM absorption nor instrumental convolution. Within the shaded region, the stellar flux is entirely absorbed by the ISM. \textit{Bottom panel :} Theoretical Lyman-$\alpha$ line profiles as they would be observed with STIS in Visit C, outside of the transit (black), and during the third orbit at mid-transit with absorption by GJ\,3470b (magenta), its exosphere alone (blue), and its ellipsoidal atmosphere alone (red). There is flux in the full-absorption ISM region because it has been spread by STIS broad LSF.}
\label{fig:Compa_absorbeurs}
\end{figure*}

Our simulations suggest that the blueshifted and redshifted absorption signatures from GJ\,3470b probe different regions of its upper atmosphere (Fig.~\ref{fig:Compa_absorbeurs}).

The in-transit to post-transit redshift of the absorption signatures measured in the red wing of the Lyman-$\alpha$ line (Sect.~\ref{sec:tr_ana}) is consistent with gas in the upper atmosphere of GJ\,3470b moving with its eccentric orbital motion (Table~\ref{tab:param_sys}; the planet radial velocity increases from $\sim$0 to 20\,km\,s$^{-1}$ between about -0.7 and 1.6\,h from mid-transit). Because interstellar neutral hydrogen absorbs the stellar flux completely within about -58 to 43\,km\,s$^{-1}$ we cannot access the saturated core from the atmospheric absorption profile, and part of its damping blue wing redshifted by the orbital motion. The rest of the damping blue wing contributes little to the observable absorption because it is located in the blue wing of the stellar Lyman-$\alpha$ line made fainter by interstellar deuterium absorption. Conversely, the orbital-motion redshift moves the damping red wing of the atmospheric absorption profile into the bright red wing of the observed Lyman-$\alpha$ line, allowing us to measure its signature with a high significance.

Meanwhile, the radiative blow-out of escaping neutral hydrogen atoms combined with the orbital velocity of the planet at the time they escape leads to a blueshifted radial velocity field in the exosphere. As a result exospheric neutral hydrogen atoms do not contribute to absorption in the observable red wing, and their velocity field is consistent with the spectral range of the absorption measured in the blue wing. Absorption from the exosphere dominates the blueshifted signature during and after the planetary transit.

\subsubsection{Red wing: an elongated upper atmosphere}
\label{sec:red_region}

\begin{figure}
\centering
\includegraphics[trim=0cm 0cm 0cm 0cm,clip=true,width=\columnwidth]{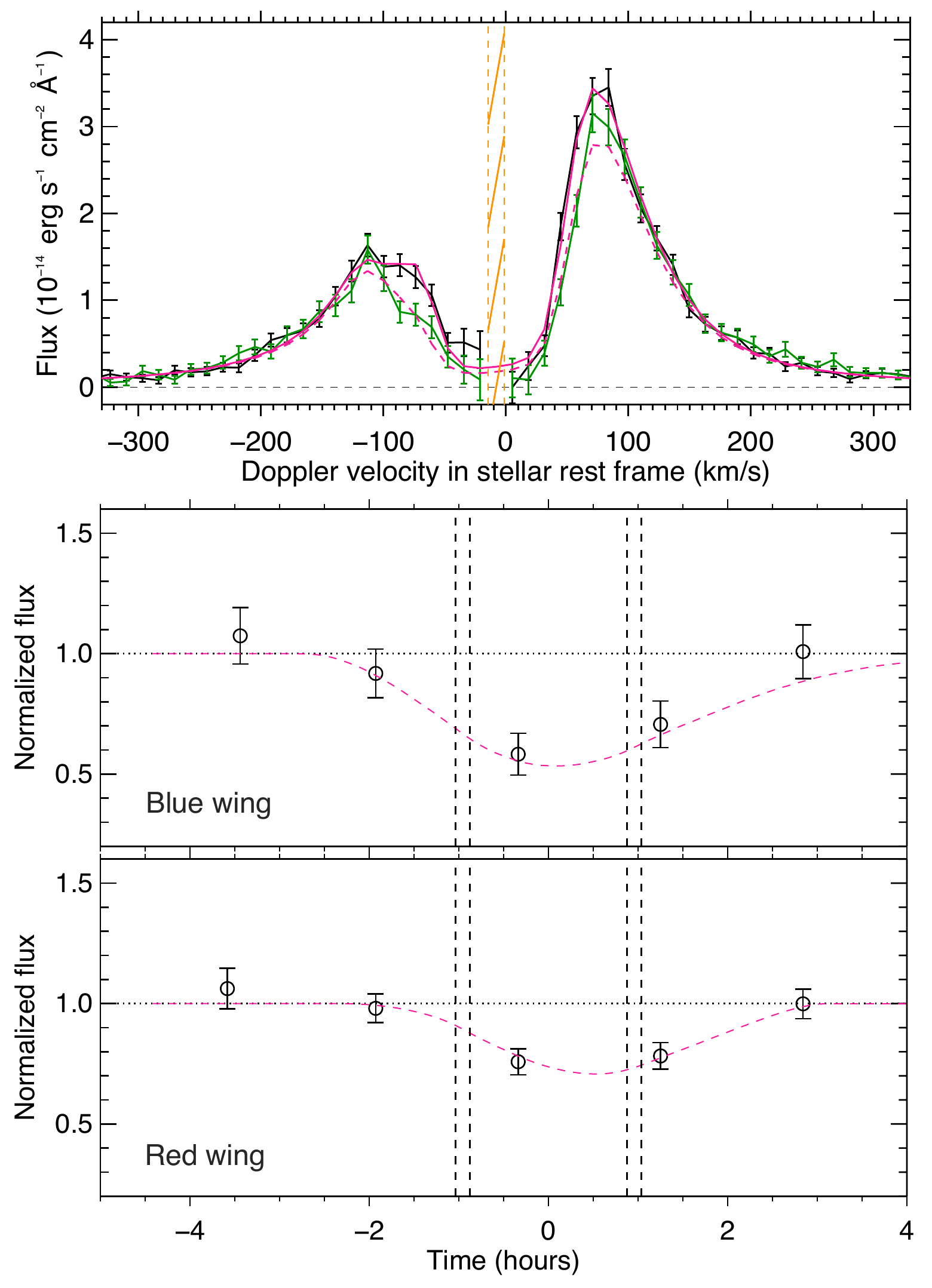}
\caption[]{Best-fit simulation to GJ\,3470 b Lyman-$\alpha$ spectra (top panel) and corresponding light curves (bottom panels). The solid magenta spectrum, which corresponds to the reconstructed intrinsic stellar spectrum, yields the dashed spectrum and light curves after absorption by the simulated upper atmosphere.}
\label{fig:Obs_sim_best}
\end{figure}

Density decreases slowly with altitude in the upper atmospheric layers because of the large temperature. The fit to the red wing signature is thus partly degenerate between the density level, the temperature, and the spatial extension of the ellipsoidal atmosphere. For example a decrease in temperature of a few thousand kelvin yields a steeper density profile, which can be compensated for by increasing the overall density of the atmosphere, by increasing its projected surface in the plane of sky, or by increasing its extension along the LOS. Interestingly though, we found that the spatial extension along the orbital motion could be constrained independently thanks to the finite duration of the Lyman-$\alpha$ transit in the red wing (Fig.~\ref{fig:Obs_sim_best}). To further investigate the properties of GJ\,3470b upper atmosphere we fixed $T_\mathrm{th}$ = 7000\,K, which is on the order of temperatures predicted by \citet{Salz2016b} for this planet. With this temperature, observations are consistent with densities of neutral hydrogen at 1\,$R_\mathrm{p}$ in between $\sim$10$^{8}$ and 10$^{10}$\,cm$^{-3}$, and hereafter $n_\mathrm{th}$ was fixed to 10$^{9}$\,cm$^{-3}$.\\

We show in Fig.~\ref{fig:Compa_absorbeurs} a best-fit model for the upper atmosphere of GJ\,3470b that reproduces well the spectral and temporal properties of the absorption signatures in the blue and the red wings (Fig.~\ref{fig:Obs_sim_best}). It yields a $\chi^2$ of 569 for 594 datapoints. The best-fit ellipsoidal atmosphere needs to be offset by $d_{||}$ = -4.5\,$\pm$2.5\,$R_\mathrm{p}$, with a radius $R_{||}$ = 22$\stackrel{+4}{_{-3}}$\,$R_\mathrm{p}$. These values do not depend on the density and temperature in the atmosphere, or on its extension in other dimensions, and suggest that neutral hydrogen gas could be present as far as $\sim$18\,$R_\mathrm{p}$ ahead of the planet, and $\sim$27\,$R_\mathrm{p}$ behind it. The radii of the atmosphere in the plane perpendicular to the orbital motion are correlated with each other because of their similar influence on the column density, and for the chosen atmospheric temperature and reference density we derive R$_{\perp}^\mathrm{orb}$ = 4$\stackrel{+3}{_{-2}}$\,$R_\mathrm{p}$ and R$_{\perp}^\mathrm{norm}$ = 7$\stackrel{+4}{_{-2}}$\,$R_\mathrm{p}$. The dense envelope of neutral hydrogen surrounding GJ\,3470b must therefore deviate significantly from a spherical shape to explain the observations. For comparison, the best-fit for a centered, spherical atmosphere yields a $\chi^2$ of 580 for a radius of 8\,$R_\mathrm{p}$. The corresponding value for the Akaike information criterion is larger by 5 points compared to the fit with an ellipsoidal atmosphere, showing that a spherical atmosphere is less than 10\% as probable to explain the observations. \\


\subsubsection{Blue wing: a blown-away exosphere}
\label{sec:blue_region}

In the frame of our simulations, Fig.~\ref{fig:Compa_absorbeurs} shows that absorption in the blue wing of the Lyman-$\alpha$ line arises from neutral hydrogen atoms that survived stellar photoionization long enough to be accelerated by radiation pressure to high velocities. Because of the strong radiative acceleration (see Sect.~\ref{sec:prad_dyn} and \citealt{Bourrier_lecav2013}), the structure of the outermost exospheric layers is not much sensitive to the altitude and velocity at which the neutral hydrogen atoms originally escaped. The density of high-velocity atoms in these layers, however, depends directly on the mass loss rate of neutral hydrogen $\dot{M}_{H^{0}}$. With radiation pressure and photoionization known, $\dot{M}_{H^{0}}$ is thus tightly constrained by the observations to a value of about 1.5$\times$10$^{10}$\,g\,s$^{-1}$, independent of the other model parameters. We note that even if the exosphere was a mix of planetary atoms and neutralized stellar wind protons (e.g. \citealt{Kislyakova2014}; \citealt{Bourrier2016}), each proton would have taken the charge of a planetary atom, and the derived mass loss rate would still trace atoms originally escaped from the planet. \\

We used the XUV spectrum derived between 5 and 911.8\,\AA\, (Sect.~\ref{sec:em_lines}) to estimate the total energy-limited mass loss rate $\dot{M}^{tot}$ from GJ\,3470 b upper atmosphere (see e.g. \citealt{Watson1981}; \citealt{Lecav2007}), accounting for the effects of tides (\citealt{Erkaev2007}). We obtain $\dot{M}^{tot}$ = $\eta$\,8.5$\times$10$^{10}$\,g\,s$^{-1}$, where $\eta$ represents the fraction of the energy input that is available for atmospheric heating. This total mass-loss rate accounts for all escaping elements, including neutral hydrogen whose loss rate can be defined as $\dot{M}_{H^{0}}$ = $f_{H^{0}}$\,$\dot{M}^{tot}$, where $f_{H^{0}}$ is the fraction of neutral hydrogen in the outflow (see e.g. \citealt{Bourrier2016}). Comparing $\dot{M}^{tot}$ with the value for $\dot{M}_{H^{0}}$ derived from the observations we obtain $\eta$\,$f_{H^{0}} \sim $ 18\%, showing that GJ\,3470b is subjected to a strong evaporation. Typical theoretical estimations of $\eta$ range between 10 and 30\% (e.g. \citealt{Lammer2013}; \citealt{Shematovich2014}; \citealt{Owen2016}), suggesting that $f_{H^{0}}$ is larger than 50\%.\\

\section{Discussion}
\label{sec:discuss}

\subsection{Possible origin for GJ\,3470b extended envelope}
\label{sec:discuss_1}

The dimensions derived for the ellipsoidal atmosphere of GJ\,3470b are dependent on our assumption of an isotropic density profile. Nonetheless, there must be dense layers of neutral hydrogen gas extending far beyond the gravitational influence of the planet to explain the observations. Indeed, the Roche lobe of GJ\,3470b extends up to 5.1\,$R_\mathrm{p}$ along the star-planet axis and has an equivalent volumetric radius of 3.6\,$R_\mathrm{p}$. Even filled with neutral hydrogen and fully opaque, it would yield a maximum absorption of $\sim$8\%, much lower than the measured redshifted absorption. While damping wings from an ellipsoidal thermosphere explain well this signature, the atmosphere of an exoplanet cannot keep expanding beyond the Roche Lobe without being sheared by stellar gravity and inertial forces. In fact, our Lyman-$\alpha$ observations only probe column densities of neutral hydrogen, and therefore we cannot constrain the exact position of the gas along the LOS. For example, we would obtain a similar fit to the data if all neutral hydrogen along the line-of-sights crossing the best-fit ellipsoidal atmosphere was concentrated in a denser shell facing the star, with the same projected surface in the plane of sky. Interestingly, such a structure could arise from the interaction between the stellar and the planetary winds. Using 2D hydrodynamical simulations, \citet{Tremblin2012} simulated the mixing layer formed by the colliding winds. The planetary outflow encounters a bow-shock at altitudes of several planet radii on the starward side, and then forms a thick and dense layer extending over tens of planetari radii along the planet motion. Similarly, 3D magnetohydrodynamic simulations performed by, e.g., \citet{Bisikalo2013}, \citet{Matsakos2015}, or \citet{Khodachenko2017} show that a strong and dense planetary outflow can move to large altitudes beyond the Roche lobe before intercepting the stellar wind. Part of the shocked, ionized planetary material can then accrete onto the star, while another part is swept back toward the planet and remains confined within its vicinity. Our observations would be sensitive to neutral hydrogen in this confined region, which would be a mix of planetary atoms and stellar wind protons neutralized via charge-exchange (\citealt{Tremblin2012}). \\

Interestingly, other exoplanets with extended atmospheres of neutral hydrogen have shown tentative redshifted absorption signatures. The evaporating hot Jupiters HD\,209458b and HD\,189733b revealed absorption signatures in between 54--187\,km\,s$^{-1}$ (5.2$\pm$1.0\%) and 60--110\,km\,s$^{-1}$ (5.5$\pm$2.7\%), respectively (\citealt{VM2003,VM2008}; \citealt{Lecav2012}; \citealt{Bourrier2013}). These signatures occured during the planetary transits, and damping wings from an extended thermosphere have been proposed as their origin in the case of HD\,209458b (\citealt{Tian2005}). The warm Neptune GJ\,436 also showed flux decreases of 15$\pm$2\% between 30--110\,km\,s$^{-1}$. A colliding wind scenario is unlikely to explain these redshifted signatures because they were observed between 0.5 and 6.6\,h after the planetary transit (\citealt{Lavie2017}) and the planetary wind has a low density favoring charge-exchange in a collisionless exosphere (\citealt{Bourrier2016}). They could rather trace an accreting stream toward the star (\citealt{Matsakos2015}; \citealt{Shaikhislamov2016}; \citealt{Khodachenko2017}; \citealt{Lavie2017}). \\

\subsection{Atmospheric escape from GJ\,3470b}
\label{sec:discuss_2}

\citet{Salz2016b} performed 1D, spherically symmetric simulations of GJ\,3470 b expanding thermosphere. They derived a heating efficiency on the order of 13\%, and predicted that the thermosphere should contain a large fraction of neutral hydrogen, down to $\sim$50\% at 4\,$R_\mathrm{p}$ and $\sim$20\% at 8\,$R_\mathrm{p}$. Our simulations also suggest that large amounts of neutral hydrogen are present in the upper atmosphere of GJ\,3470b, albeit at higher altitudes than in the simulations by \citet{Salz2016b}. Their results depend on the XUV and Lyman-$\alpha$ stellar irradiation, which they assumed to be about twice and four times larger than our observationnaly-derived values. A lower irradiation in their simulations would increase the fraction of neutral hydrogen at a given altitude, which would be more consistent with our observations. \\

Our estimate for the intrinsic stellar Lyman-$\alpha$ (1.85\,erg\,cm$^{-2}$\,s$^{-1}$, after scaling to a stellar radius of 0.39\,$R_\odot$) and X-ray (0.82\,erg\,cm$^{-2}$\,s$^{-1}$) fluxes yields ages $\tau$ = 2.1 and 2.2\,Gyr from the M dwarfs relations in \citet{Guinan2016}, respectively. The rotation period of GJ\,3470 (20.7\,days) further yield $\tau$ = 1.9\,Gyr from the relation for early-type M dwarfs in \citet{Engle2018}. The consistency between the derived ages strengthens the reliability of these various relations. With total mass loss rate ranging between the neutral hydrogen mass loss rate ($\sim$1.5$\times$10$^{10}$\,g\,s$^{-1}$) and the maximum energy-limited mass loss rate (8.5$\times$10$^{10}$\,g\,s$^{-1}$), GJ\,3470b would have lost between 1.1 and 6.5\% of its current mass over its lifetime of $\sim$2\,Gyr. Because of a larger stellar XUV emission at the beginning of the star's life (e.g. \citealt{Shkolnik2014}), the total atmospheric escape over the planet lifetime is likely about 5 times stronger (Eq 19 in \cite{Lecav2007}), resulting in a mass loss between $\sim$4 and 35\% of the total planet mass. A similar conclusion is reached when comparing the cumulative XUV irradiation and escape velocity from GJ\,3470b with that of Solar system and extrasolar planets, as proposed by \citet{Zahnle2017}\footnote{We use the authors' conventions to place GJ\,3470b in their Figure 2, with an XUV irradiation of 230 times that of the Earth, and an escape velocity of 19\,km\,s$^{-1}$ }. GJ\,3470b is located above their cosmic shoreline, \textit{i.e.} in a region where exoplanets might have lost a significant fraction of their atmosphere, and is predicted to have evaporated about 20\% of its mass (assuming $\eta$=20\%).  \\

We can compare the upper atmospheric properties of GJ\,3470b with those of GJ\,436b, which is subjected to an XUV irradiation about 3 times lower (3938 vs 1303\,erg\,cm$^{-2}$\,s$^{-1}$, see Sect.~\ref{sec:em_lines} and \citealt{Bourrier2016}). \citet{Bourrier2016} constrained $\eta$\,$f_{H^{0}}$ to be on the order of 10$^{-2}$ for GJ\,436b, with a lower limit of 0.5\% on both $\eta$ and $f_{H^{0}}$. A larger irradiation, combined with a lower heating efficiency and/or higher ionization fraction, likely explains why GJ\,3470b loses today about a hundred times more neutral hydrogen than GJ\,436b ($\sim$2.5$\times$10$^8$\,g\,s$^{-1}$). The mass loss rate from GJ\,3470b is more in the range of theoretical and derived values for hot Jupiters like HD\,209458 (e.g. \citealt{Bourrier_lecav2013}), which is actually subjected to a lower XUV irradiation (2929\,erg\,cm$^{-2}$\,s$^{-1}$, \citealt{Louden2017}).\\

\subsection{Influence of radiation pressure and stellar wind}
\label{sec:discuss_3}

\label{sec:prad_dyn}

\begin{figure}
\centering
\includegraphics[trim=0cm 0cm 0cm 0cm,clip=true,width=\columnwidth]{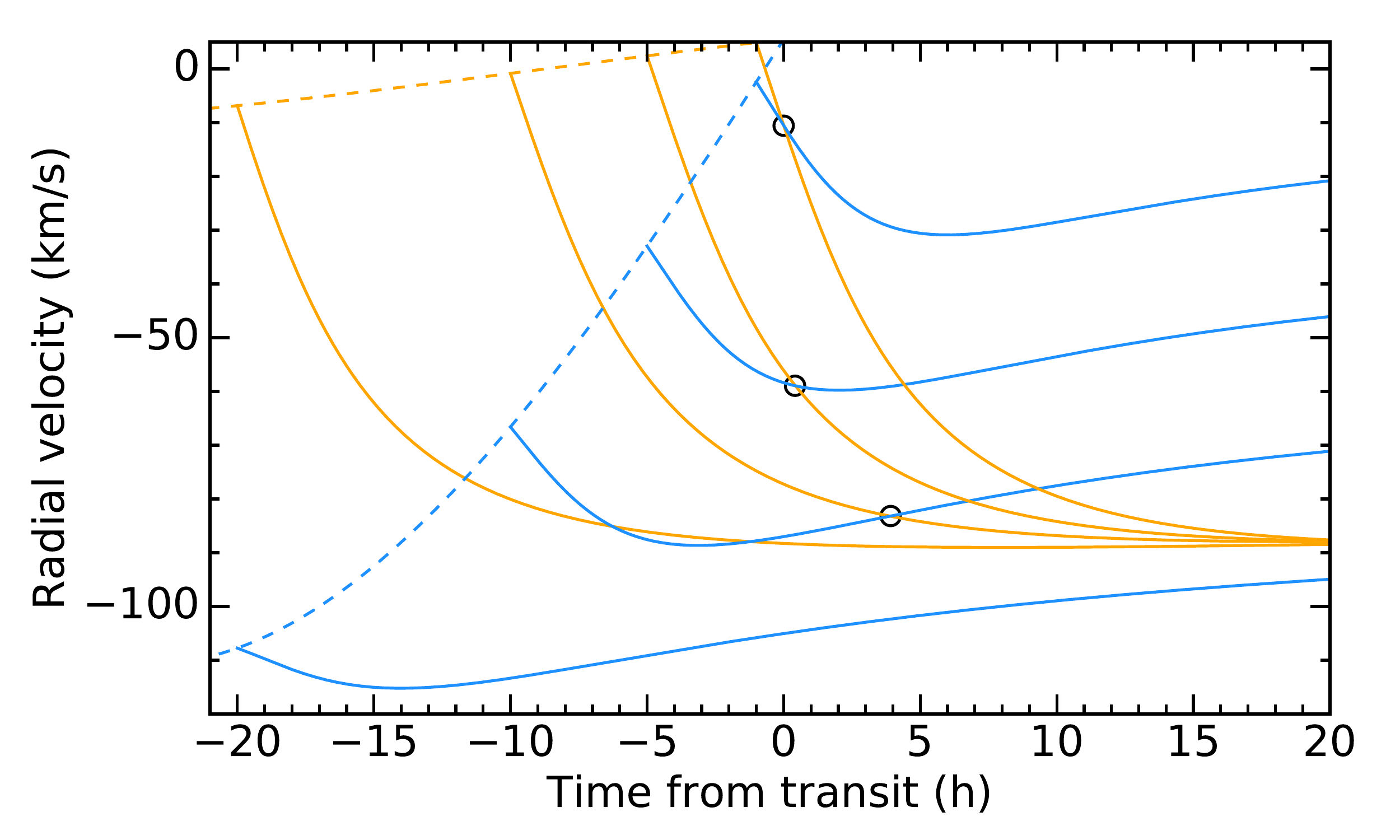}
\caption[]{Radial velocity of a neutral hydrogen atom escaping from GJ\,3470 b as a function of time, projected on the star-atom axis (solid orange line) and on the LOS (solid blue line). The different curves correspond to four times of escape, which can be read at the intersection with the planet velocity curves (dashed lines). Black circles highlight the time of transit of the atoms.}
\label{fig:GJ3470b_RVcurves}
\end{figure}

In the present configuration of radiative blow-out (\citealt{Bourrier_lecav2013}), the velocity of exospheric atoms in the star rest frame is the combination of the orbital velocity of the planet at the time they escaped its gravitational influence (tangential to the orbital trajectory), and the velocity they then acquired under radiation pressure acceleration (in the opposite direction to the star). The velocity of a transiting atom projected on the LOS then determines the Doppler velocity of its absorption signature. The radial velocity of GJ\,3470b increases from -110\,km\,s$^{-1}$ near the approaching quadrature to about -5 to 15\,km\,s$^{-1}$ during transit. Thus, the planet has a larger contribution on the radial velocity of particles escaping long before the transit. Furthermore, radiation pressure induces a similar dynamics on all escaping hydrogen atoms (neglecting self-shielding effects and the eccentricity of GJ\,3470b orbit) but it takes up to 20\,h to accelerate them up to their terminal velocity (Fig.~\ref{fig:GJ3470b_RVcurves}). Therefore, the radial velocity of a hydrogen atom at the time it transits the star depends on when it escaped the planet. Absorption is measured up to about -90\,km\,s$^{-1}$, and simulations predict a peak of absorption at about -60\,km\,s$^{-1}$ (Fig.~\ref{fig:Compa_absorbeurs}). Fig.~\ref{fig:GJ3470b_RVcurves} shows that neutral hydrogen atoms must escape the planet more than 5\,h before its transit to occult the star with radial velocities larger than -60\,km\,s$^{-1}$. However, the short photoionization lifetime ($\sim$1\,h, Sect.~\ref{sec:em_lines}) implies that a large fraction of the neutral hydrogen atoms escaping from GJ\,3470b are ionized before they reach the observed velocities. This is why neutral hydrogen must escape in such large amounts from GJ\,3470b to yield the absorption measured at high velocities in the blue wing. \\

The uncertainties on the reconstructed Lyman-$\alpha$ line profile only affect the radial velocity of neutral hydrogen atoms by a few km\,s$^{-1}$ during the transit, and thus have no significant impact on the simulations and our conclusions. Recent hydrodynamical simulations suggest that the outflow from the hot Jupiter HD\,209458b is strongly collisional beyond the Roche Lobe, preventing radiation pressure from imparting momentum to escaping neutral hydrogen atoms (\citealt{Khodachenko2017}, \citealt{Cherenkov2018}). Direct observations of the outflow at lower altitudes in the expanding thermosphere are required to assess the validity of this scenario (\citealt{Spake2018}, \citealt{Oklopcic2018}). Interestingly though, all Lyman-$\alpha$ transit observations are well reproduced with a model of a collision-less exosphere accelerated by radiation pressure (\citealt{VM2003}, \citealt{Ehrenreich2012}, \citealt{Bourrier_lecav2013}), combined in some cases with charge-exchange with the stellar wind (\citealt{Lecav2012}, \citealt{Bourrier_lecav2013}, \citealt{Bourrier2016}, \citealt{Lavie2017}). Our present study of GJ\,3470b shows that the blueshifted component of its Lyman-$\alpha$ absorption signature can be explained by a radiatively blown-out exosphere, without needing to invoke interactions with the stellar wind (Fig.~\ref{fig:Compa_absorbeurs}). This contrasts with the planet GJ\,436b, whose exosphere is shaped by the combination of radiative braking and charge-exchange with the low-velocity ($\sim$55\,km\,s$^{-1}$) wind of its host star (\citealt{Bourrier2016}). Since GJ\,3470 is an earlier-type, younger M dwarf than GJ\,436 (which has an age greater than 4\,Gyr, \citealt{Bourrier_2018_Nat}, \citealt{Veyette2018}), its stellar wind might be faster than the observable range of the stellar Lyman-$\alpha$ line ($\sim$200 km/s), which would explain why stellar wind protons neutralized by charge-exchange with the exosphere absorb too far in the wing to be detectable. Nonetheless the wind from GJ\,3470 could still have an influence on the upper atmosphere of its planet, for example via the formation of a mixing layer (see Sect~\ref{sec:red_region}). \\

\section{Summary of findings and perspectives}
\label{sec:conclu}

We observed the transit of the warm Neptune GJ\,3470b in the Lyman-$\alpha$ line at three different epochs, and obtained the following results:
\begin{enumerate}
\item We measure absorption depths of 35$\pm$7\% in the blue wing of the line, between -94 and -41\,km\,s$^{-1}$, and of 23$\pm$5\% in the red wing, between 23 and 76\,km\,s$^{-1}$.

\item These absorption signatures are repeatable over the three epochs and phased with the planet transit.

\item The redshifted absorption requires high densities of neutral hydrogen in the vicinity of the planet, in an elongated region extended up to $\sim$18\,$R_\mathrm{p}$ ahead of the planet and $\sim$27\,$R_\mathrm{p}$ behind it.

\item The blueshifted absorption is well explained by neutral hydrogen atoms escaping at a rate of $\sim$1.5$\times$10$^{10}$\,g\,s$^{-1}$ and blown away from the star by radiation pressure. This may have led to the loss of 4--35\% of GJ\,3470b current mass over its $\sim$2\,Gyr lifetime.

\end{enumerate}
GJ\,3470b is the second warm Neptune after GJ\,436b to be observed evaporating and surrounded by an extended upper atmosphere of neutral hydrogen, supporting the idea that this class of exoplanets are ideal targets to be characterized through their upper atmosphere. GJ\,3470b shows a larger mass loss and a smaller exosphere than GJ 436b. This likely results from the lower density of GJ\,3470b, the stronger stellar energy input, photoionization, and radiation pressure from GJ\,3470, and possibly different wind properties. Observations of other escaping species not hindered by ISM absorption, such as helium at infrared wavelengths (\citealt{Oklopcic2018}, \citealt{Spake2018}), would allow us to probe low-velocity regions in the upper atmosphere of these warm Neptunes to better understand their structure and formation. Further studies of these planets’ present and past atmospheric properties are required to understand their presence at the edge of the evaporation desert, the role of their M dwarf host stars, and the influence of their past dynamical evolution (see \citealt{Bourrier_2018_Nat} for GJ\,436b).


\begin{acknowledgements}
We thank the referee for enlightening comments. We offer our thanks to Nathan Hara for his help with the random sampling of an ellipsoidal surface, and to Alexandre Correia for discussions about thermal tides. The research leading to these results received funding from the European Research Council under the European Union's Seventh Framework Program (FP7/2007-2013)/ERC grant agreement number 336792. This work is based on observations made with the NASA/ESA HST (PanCET program, GO 14767), obtained at the Space Telescope Science Institute (STScI) operated by AURA, Inc. Support for this work was provided by NASA grants under the HST-GO-14767 program of the STScI. This work has been carried out in the frame of the National Centre for Competence in Research ``PlanetS'' supported by the Swiss National Science Foundation (SNSF). R.A. acknowledges the financial support of the SNSF. This project has received funding from the European Research Council (ERC) under the European Union's Horizon 2020 research and innovation programme (project Four Aces grant agreement No 724427). A.L.E. acknowledges support from CNES and the French Agence Nationale de la Recherche (ANR), under programme ANR-12-BS05-0012 “Exo-Atmos”. J.S.-F. acknowledges support from the Spanish MINECO grant AYA2016-79425-C3-2-P. This work has made use of data from the European Space Agency (ESA) mission
{\it Gaia} (\url{https://www.cosmos.esa.int/gaia}), processed by the {\it Gaia}
Data Processing and Analysis Consortium (DPAC,
\url{https://www.cosmos.esa.int/web/gaia/dpac/consortium}). Funding for the DPAC
has been provided by national institutions, in particular the institutions
participating in the {\it Gaia} Multilateral Agreement.
\end{acknowledgements}

\bibliographystyle{aa} 
\bibliography{biblio} 

\end{document}